\documentclass[10pt,journal,twoside]{IEEEtran}

\usepackage{cite}
\usepackage{url}
\usepackage[colorlinks,linkcolor=blue,anchorcolor=green,citecolor=red, urlcolor=red, backref=false]{hyperref}

\usepackage{graphicx}
\usepackage{subfigure}
\usepackage{verbatim}
\usepackage{amsmath}
\usepackage{amsfonts}
\usepackage{amssymb}

\usepackage{algorithm}
\usepackage{algorithmic}

\usepackage{booktabs}
\usepackage{multirow}
\usepackage{multicol}

\usepackage{booktabs}

\usepackage{diagbox} %
\usepackage{amssymb} %
\usepackage{bm}

\usepackage{balance}

\usepackage{caption}

\begin{document}
\title{FedMUA: Exploring the Vulnerabilities of Federated Learning to Malicious Unlearning Attacks}

\author{Jian~Chen,~\IEEEmembership{Member,~IEEE},
~Zehui~Lin, 
~Wanyu~Lin*,~\IEEEmembership{Member,~IEEE},
~Wenlong~Shi,\\
~Xiaoyan~Yin,~\IEEEmembership{Member,~IEEE},
~Di~Wang,~\IEEEmembership{Member,~IEEE}

\IEEEcompsocitemizethanks{
\IEEEcompsocthanksitem This work was supported in part by Hong Kong
(HK) Research Grant Council (RGC) General Research Fund under Grant
PolyU 15208222, and in part by the NSFC Young Scientist Fund under Grant
PolyU A0040473. (\textit{Corresponding author: Wanyu Lin.)}
\IEEEcompsocthanksitem J. Chen and Z. Lin are with the Department of Data Science and Artificial Intelligence, The Hong Kong Polytechnic University, Hong Kong, China. Email: comp-jian.chen@polyu.edu.hk, linzehui19@gmail.com.
\IEEEcompsocthanksitem W. Lin is with the Department of Data Science and Artificial Intelligence and the Department of Computing, The Hong Kong Polytechnic University, Hong Kong, China. Email: wan-yu.lin@polyu.edu.hk.
\IEEEcompsocthanksitem W. Shi is with 
School of Electronic Information and Communications, Huazhong University of Science and Technology, Wuhan 430074, China.
Email: wenlongshi@hust.edu.cn.
\IEEEcompsocthanksitem X. Yin is with the School of Information Science and Technology, Northwest University, Xi’an, China. E-mail: SCxiaoyanyin@gmail.com.
\IEEEcompsocthanksitem D. Wang is with the Division of Computer, Electrical and Mathematical Sciences and Engineering, King Abdullah University of Science and Technology, Saudi Arabia. E-mail: di.wang@kaust.edu.sa.
}
}
\markboth{IEEE Transactions on Information Forensics and Security}
{J. Chen \MakeLowercase{\textit{et al.}}: FedMUA: Exploring the Vulnerabilities of Federated Learning to Malicious Unlearning Attacks}

\maketitle
	\begin{abstract}
		
Recently, the practical needs of ``the right to be forgotten'' in federated learning gave birth to a paradigm known as federated unlearning, which enables the server to forget personal data upon the client's removal request. Existing studies on federated unlearning have primarily focused on efficiently eliminating the influence of requested data from the client's model without retraining from scratch, however, they have rarely doubted the reliability of the global model posed by the discrepancy between its prediction performance before and after unlearning. To bridge this gap, we take the first step by introducing a novel malicious unlearning attack dubbed FedMUA, aiming to unveil potential vulnerabilities emerging from federated learning during the unlearning process. Specifically, clients may act as attackers by crafting malicious unlearning requests to manipulate the prediction behavior of the global model. The crux of FedMUA is to mislead the global model into unlearning more information associated with the influential samples for the target sample than anticipated, thus inducing adverse effects on target samples from other clients. To achieve this, we design a novel two-step method, known as Influential Sample Identification and Malicious Unlearning Generation, to identify and subsequently generate malicious feature unlearning requests within the influential samples. By doing so, we can significantly alter the predictions pertaining to the target sample by initiating the malicious feature unlearning requests, leading to the deliberate manipulation for the user adversely. Additionally, we design a new defense mechanism that is highly resilient against malicious unlearning attacks. Extensive experiments on three realistic datasets reveal that FedMUA effectively induces misclassification on target samples and can achieve an 80\% attack success rate by triggering only 0.3\% malicious unlearning requests.

	\end{abstract}
	
	\section{Introduction}
	\label{sec:intro}
	
	Over recent years, federated learning (FL)~\cite{lyu2024secure, du2024towards,qin2023reliable, li2023revisiting,lu2023federated,yu2023multimodal} has emerged as a novel paradigm for collaborative machine learning (ML). In practice, clients may opt to entirely erase their data from the trained FL models for many reasons, including privacy, usability, and data fidelity. Recent legislations such as the General Data Protection Regulation (GDPR)~\cite{voigt2017eu} and the California Consumer Privacy Act (CCPA)~\cite{harding2019understanding} empower clients with ``the right to be forgotten'', i.e., the right to have their data erased from a well-built FL system upon removal requests. Consequently, this has given rise to a new paradigm known as federated unlearning (FU), which enables the server to effectively forget personal data upon the client's removal request.

	\begin{figure}[t]
		\centering
		\includegraphics[scale=0.47]{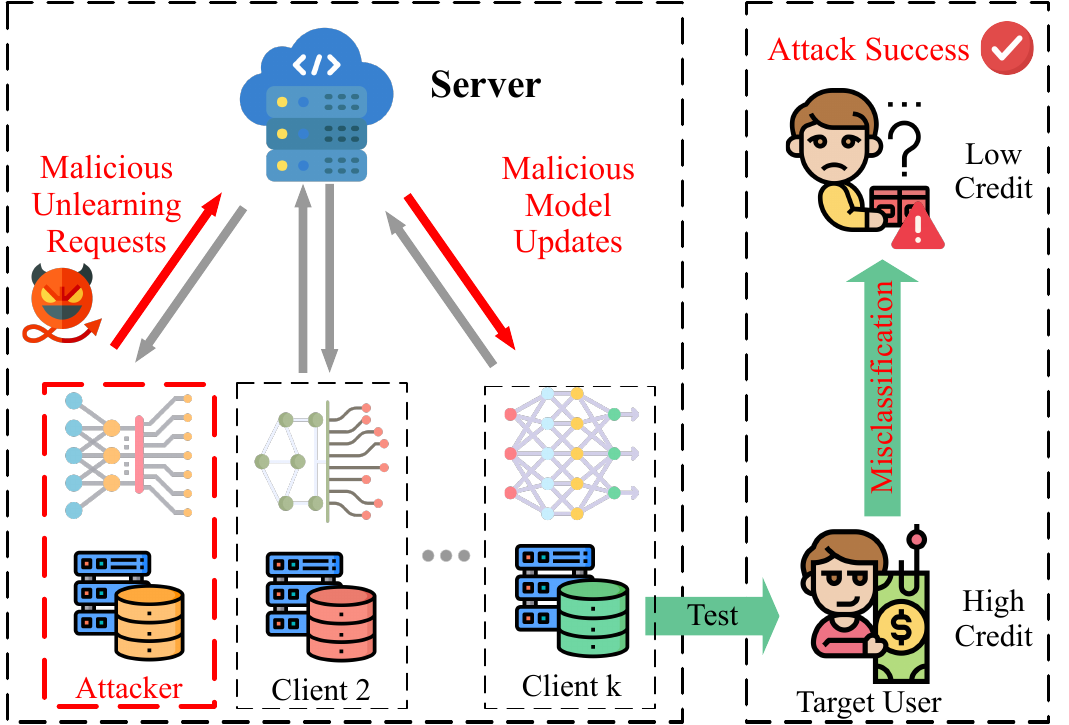}
		\caption{Overview of FedMUA. Taking credit rating as an example, several clients collaboratively train a global model. A particular client acts as an attacker, aiming to submit malicious unlearning requests to the server. The server will receive malicious unlearning gradient updates from this client. After updating these gradients on the global model and each local model, the target user initially identified with a high credit rating from other clients is successfully misclassified as having a low credit rating.
		}
		\label{fig:overview}
	\end{figure}

	Existing studies~\cite{gao2024verifi, liu2021federaser, su2023asynchronous, che2023fast, zhang2023fedrecovery} on federated unlearning primarily focus on developing effective methods for erasing a client's data. However, existing works do not consider the security vulnerabilities that could arise in the global model after the unlearning process. To fill this gap, in this paper, we take the initial step in investigating how federated unlearning can be leveraged maliciously to potentially manipulate the FL models. We thus introduce a novel malicious unlearning attack for FL, termed FedMUA, which deliberately leads to adverse prediction behaviors on the target data. The intentional use of FedMUA to manipulate predictions could potentially raise ethical concerns in security-critical applications. For instance, in credit rating, when several competitive banks collaborate to build a credit rating model jointly (c.f. Fig.~\ref{fig:overview}), a particular client may act as an attacker to initiate malicious unlearning requests to the server to assign low credit ratings for other client's users, causing significant inconvenience to the user's life.
	
	FedMUA is founded on the intuition that unlearning useful information about the target sample from the model can lead to a significant alteration in its predictions. To induce substantial prediction bias, it is essential to identify and eliminate the influential samples within the training dataset from the attacker that exerts the most substantial influence on predictions for the target sample. However, directly unlearning the influential samples is ineffective mainly due to the inherent distinction in the way how FL and traditional ML learn from data, as well as the aggregation impact of each client's data on the target sample. 
	
	Given the aforementioned challenge, an intuitive approach is to modify the labels of these influential samples. However, this kind of modification can be easily detected by the server. To mimic the normal users’ unlearning request behaviors and measure the feasibility of potential risks, we resort to generating malicious feature unlearning requests for these influential samples. Particularly, the global model unlearns information beyond the intended scope by manipulating the normal unlearning requests of influential samples, containing more knowledge than anticipated. Consequently, when the server unlearns these malicious requests, the global model will remove additional information about the target sample, ultimately influencing its predictions.

	Considering the above ideas, we introduce \textit{Influential Sample Identification (ISI)} as an efficient method for identifying influential samples for a given target sample. Specifically, we leverage influence functions (IFs)~\cite{koh2017understanding} to identify the influential samples, which are commonly used to assess the impact of each training sample on the prediction of specific data. To ensure the attack's effectiveness, we prioritize selecting influential samples with negative IF values and strive to obtain the identical label for them. We further propose \textit{Malicious Unlearning Generation (MUG)} to generate malicious feature unlearning requests associated with these influential samples by pushing the unlearned features of them to be closer to the features of the target data. By doing so, we can substantially introduce bias into predictions pertaining to the target sample, yielding the deliberate manipulation for the target user adversely.

    To mitigate such attacks, it is crucial to identify malicious unlearning requests and reduce their negative impact on the global model. To date, no effective defense mechanisms have been designed to counter these attacks. Existing studies on the security vulnerabilities of FL mainly focus on adversarial attacks and data poisoning attacks, failing to address the failure modes of FL caused by malicious unlearning attacks. To alleviate the negative impact caused by malicious unlearning requests, we visualize the gradient updates for each client in each round using three realistic datasets and make a key observation: the gradient updates to the server by malicious clients are much larger than those of normal clients in the initial training rounds. On this basis, the identification of malicious unlearning requests can be facilitated by focusing on the larger gradient updates. Subsequently, the server can reduce the malicious gradients to mitigate such attacks.

	We summarize our major contributions as follows:
	\begin{itemize}
		
		\item We introduce FedMUA, the first work to unveil previously unexplored threats in real-world federated unlearning pipeline. This effort identifies the potential risk of intentionally manipulating the predictive behavior of the global model for the target sample against FL systems .
		
		\item  We propose a novel two-step attack method, namely ISI and MUG that allows a client to achieve FedMUA with only black-box access to the global model, rendering it easily applicable in real-world federated unlearning settings.
		
	    \item  We design a new defense mechanism, the first to provide high resilience against malicious unlearning attacks in federated unlearning settings. This is achieved by identifying malicious unlearning requests and reducing their negative impact based on observations from realistic datasets.

		\item We empirically evaluate FedMUA on multiple datasets, FU methods, and aggregation rules. The results demonstrate that the existing federated unlearning frameworks lack robustness against FedMUA. The code of FedMUA has been released for reproducibility purposes\footnote{\url{https://github.com/ity207/FedMUA}}.

	\end{itemize}
	
	\section{Related Works}
	\label{sec:Related Works}
	
	\subsection{Federated Unlearning}
	
	Federated unlearning is an emerging paradigm that effectively updates a FL model after completely removing data without requiring it to be retrained from scratch. Existing federated unlearning studies~\cite{gao2022verifi,che2023fast} mainly concentrate on designing effective methods to erase a client's data. For instance, Liu et al.~\cite{liu2021federaser} take the first step by proposing FedEraser, the first federated unlearning methodology that can eliminate the influence of a client's data on the global FL model while significantly reducing the time used for constructing the unlearned FL model. Wu et al.~\cite{wu2022federated} further eliminate a client's contribution by subtracting the accumulated historical updates from the model and leveraging the knowledge distillation technique to restore the model's performance without relying on any data from the clients. In addition, Wang et al.~\cite{wang2022federated} observe that different channels have a varying contribution to different categories in image classification tasks and then propose a method for removing the information about particular categories. Liu et al.~\cite{liu2022right} propose a smart retraining method for federated unlearning without communication protocols and Su et al.~\cite{su2023asynchronous} introduce KNOT, a clustered aggregation mechanism designed for asynchronous federated learning. In contrast to the current federated unlearning techniques which mainly focus on erasing the information of client's data effectively, our work concentrates on the inherent security vulnerability in the unlearning process.

	\subsection{Unlearning Attacks}
	
	Unlearning attacks are a category of attacks that take place during the unlearning process. So far, unlearning attacks have primarily been studied in the context of machine learning models. For instance, Di et al.~\cite{di2022hidden} conduct their attack by introducing both poisoned and camouflage sets into the training dataset. This strategy aimed to ensure that the model's prediction behaviors closely resembled those of the original data. Subsequently, an unlearning request is triggered to remove the camouflage dataset, which initiated the activation of the poison effect within the model. Qian et al.~\cite{qian2023towards} further uncover the vulnerability of deep neural networks (DNN) during the unlearning process. They generate malicious unlearning requests by identifying features capable of causing misclassification in the target data. Zhao et al.~\cite{zhao2024static} explore static and sequential malicious attacks in the context of selective forgetting, formulating them as a stochastic optimal control problem. By targeting critical data update requests, they aim to maximize the impact of malicious actions. Additionally, Hu et al.~\cite{hu2023duty} craft malicious unlearning requests in the context of Machine Learning as a Service (MLaaS) scenario. By incorporating samples from a different task into the original unlearned sample, the overall performance of the model will be heavily compromised. Building on insights from these previous works, we have identified that initiating malicious unlearning requests can serve as a potential approach for launching targeted attacks on federated learning models. The unlearning process, which is designed to remove the information of the data from the model, can be exploited if manipulated maliciously. By understanding these vulnerabilities, we aim to enhance the robustness of federated learning systems.

	\subsection{Influence Functions}
	
	Influence Functions serve as a valuable tool with a wide range of applications, including reassigning labels to harmful training data~\cite{kong2021resolving}, fairness in models' predictions~\cite{sun2023regularizing}, and providing insights into model predictions~\cite{saunshi2023understanding,tsai2023representer}. Specifically, Koh and Liang~\cite{koh2017understanding} use the implicit Hessian-vector product to compute Influence Functions for training data. Chen et al.~\cite{chen2021hydra} extend this approach by investigating the training trajectory and enhancing computational efficiency, eliminating the need for Hessian matrix computations. Additionally, influence function-based methods can be computationally expensive, primarily due to the calculation of the Hessian, especially in large-scale federated learning systems. To address the run-time overhead, subsequent studies have focused on accelerating the calculation time. For instance, Guo et al.~\cite{guo2021fastif} introduce Fast Influence Functions (FastIF), which incorporate various speed-up heuristics, while Schioppa et al.~\cite{schioppa2022scaling} employ Arnoldi’s iteration algorithm to efficiently identify the dominant eigenvalues and eigenvectors of the inverse Hessian. By integrating these improvements, our work will become more practical.  In our work, we leverage Influence Functions to pinpoint influential samples and enhance the generation of malicious unlearning for such samples.

	\section{Threat Model}
	\label{sec:threat}

	In this section, we outline the scenario we follow and clarify our underlying assumptions. Subsequently, we provide details about the attacker's goal, knowledge, and capabilities.
	
	\subsection{Attacker's Goal}
	Following the training process of FL, we make the assumption that $K$ clients (e.g., several competitive banks) possess data that can be either IID (independently and identically distributed) or Non-IID in nature. These clients jointly agree on a unified learning goal and collaboratively train a global credit rating model for the server. Under this setting, we consider one or more untrusted clients may act as the possible attackers, who aim to penalize a particular target user from other clients due to some potential conflicts of interests. By initiating malicious unlearning requests, the global model can be manipulated to incorrectly classify a target user, who initially had a high credit rating, as having a low rating. This misclassification not only inconveniences the target user but also diminishes the reputation of other clients.

	In this scenario, the attacker's goal is to manipulate the unlearning process in a way that the resulting federated unlearned model adversely affects the target user. Simultaneously, it is expected that this unlearned model will maintain its prediction performance for non-target users from other clients, ensuring the stealthy and inconspicuous nature of our attack. More specifically, we can define the attacker's goal as the target effectiveness goal and the model utility goal as follows:
	
	\begin{itemize}
		\item \textbf{Goal I: target effectiveness goal.} This goal implies the deliberate misclassification behavior on the particular target user by the malicious unlearned local client model (i.e., the malicious unlearned global model) during the test-time phase. It aims to guarantee that the malicious unlearned local client model can predict the desired prediction for the target user.
		
		\item \textbf{Goal II: model utility goal.} This goal refers to the unaffected prediction performance of the malicious unlearned local client model on non-target users. It is intended that the malicious unlearned client model can make the same predictions as the local client model trained without any unlearning requests for non-target users.
	\end{itemize}

	\subsection{Attacker's Knowledge}
	Here, we consider a realistic attack model in real-world applications and delve into the black-box settings. More precisely, we assume that the attacker does not obtain any prior knowledge about the global model and the local models of other clients, including the model architecture and model parameters. In our scenario, the attacker is assumed to obtain the following knowledge. First, the attacker knows which user from other clients he intends to attack (e.g., a particular user with a high credit rating in the context of the credit rating task). Second, the attacker has the capability to manipulate the features of a small ratio of the data in the local client's training data and send unlearning requests to the server. Note that this is reasonable in practice where the client may act as an attacker. Particularly, in a credit scoring system, the race, gender or other features might be replaced with different content.
	
	Furthermore, it should be noted that since the attacker does not compromise the global training procedure, no prior knowledge about the other clients' training dataset used to train the global model is required.

	\subsection{Attacker's Capability}
	
	To initiate our attack, the attacker's capability is constrained by imposing an upper limit on the number of influential samples that can be unlearned from the local model, whose manipulated features are unlearned by the server. Moreover, the manipulated influence on the target data is limited by the quantity of attackers. The greater the number of attackers, the more effective the attack becomes.
	
	In this scenario, let the total number of local training data be denoted as $n$, and the number of influential samples be $m (\ll n)$. Consequently, the attacker's influence extends over a fraction $\alpha = m/n$ of the training data. This constraint ensures that the attacker's unlearning requests do not adversely affect predictions for non-target data.
	
	\section{Problem Formulation}
	
	\label{sec:problem}
	
	We focus on malicious unlearning attacks against FL models. In this context, we consider a FL system with $K$ local clients. Each local client $i$ has a training dataset $\mathcal{D}_i$, where $i=1,\cdots,K$. Generally, the coordination of a central server facilitates the collaborative training of a unified model by these $K$ clients' training data. In particular, FL aims to learn a global model with its parameters $\textbf{w}$ to minimize the averaged loss as follows:
	\begin{align}
		\min_{\mathbf{w} \in\mathbb{R}^d} \frac{1}{K}  \sum_{i =1}^{K}  \mathcal{L}_i(\mathbf{w}, \mathcal{D}_i), 
		\label{FL_objective}
	\end{align}
	where $\mathcal{L}_i(\mathbf{w}, \mathcal{D}_i) = \frac{1}{| \mathcal{D}_i |} \sum_{\zeta \in \mathcal{D}_i} \mathcal{L}(\mathbf{w}, \zeta)$ is the local training loss for client $i$, $| \mathcal{D}_i|$ is the number of training data of client $i$, and $d$ is the dimension of global model $\mathbf{w}$.
	
	In the federated unlearning process, FL performs the following two fundamental steps:
	
	\begin{itemize}
		
		\item \textbf{Step I.} 
		During the FL training process, each client initially downloads the global model $\mathcal{M}_G(.,\textbf{w})$ from the central server. Subsequently, in each training round, every client independently trains its local model $\mathcal{M}_i(.,\textbf{w}_{i})$ using its own local training data $\mathcal{D}_i$. The central server then collects all the updates from the K clients and proceeds to update the global model through an aggregation of these collected updates. This results in an updated model $\mathcal{M}_{i+1}(.,\textbf{w}_{i+1})$ which serves as the global model for the subsequent training round.
		
		\item \textbf{Step II.} 
		Client $i$ sends data removal requests $\mathcal{D}_f$ at a certain time and the server aims to eliminate the influence of the data $\mathcal{D}_f$ from the global model $\mathcal{M}_G(.,\textbf{w})$. It is worth noting that in the federated learning process, the server and the clients communicate with each other using gradients, and client $i$ could send the updated model $\mathcal{M}_i(.,\textbf{w}^{'}_i)$ trained without $\mathcal{D}_f$ to the server. Then, the server could use federated unlearning algorithm to update the global model without retraining.

	\end{itemize}
	
	In our attack, we first let $\boldsymbol x_t$ be the target data and $y_t$ the corresponding label from a target client with its local model $\mathcal{M}_t(.,\textbf{w}_t)$. For malicious client $m$ with its training data (i.e., $\mathcal{D}_m=\{(\boldsymbol x_j, y_j)\}_{j=1}^m$) and the local model $\mathcal{M}_m(.,\textbf{w}_m)$, the client first finds $p$ influential training samples (denoted as $\mathcal{D}_{inf}=\{(\boldsymbol x_j, y_j)\}_{j=1}^p$) for the target data, and then makes the corresponding malicious unlearning modification (i.e., $\boldsymbol \delta_j$) on each $\boldsymbol x_j$. The malicious client then updates the local model $\mathcal{M}_m(.,\textbf{w}^{'}_m)$ trained without $\{\boldsymbol \delta_j\}_{j=1}^p$ and sends it to the server for further unlearning. The goal of FedMUA is to generate malicious feature unlearning requests (i.e., $\{\boldsymbol \delta_j\}_{j=1}^p$) and obtain the updated local model $\mathcal{M}_m(.,\textbf{w}^{'}_m)$ to update the global model $\mathcal{M}_G(.,\textbf{w})$ as follows:

	\begin{align}
		\mathcal{M}_t  (\boldsymbol x_t, \textbf{w}^{u}_t)\neq y_t,
	\end{align}
	\vspace{-4ex}
	\begin{align}
	\mathcal{M}_G(.,\textbf{w}^{u}) =U( \mathcal{M}_m(.,\textbf{w}^{'}_m), \sum_{i =1,i\neq m}^{K} \mathcal{M}_i(.,\textbf{w}_i), ),
	\end{align}
	\vspace{-4ex}
	\begin{align}
		\lVert\boldsymbol \delta_j \rVert \leq \epsilon,
	\end{align}
	where $U(\cdot)$ is the unlearning algorithm, $\epsilon$ is the maximal feature modifications of the requested data, $\textbf{w}^{u}_t$ is the unlearned targe model, $x_t$ is the target data, and $\mathcal{M}_t  (., \textbf{w}^{u}_t)$ represents the unlearned model for the target client.
	
	\section{Attack Methodology}
	\label{sec:Methodology}

	\begin{figure}[t]
		\centering
		\includegraphics[scale=0.45]{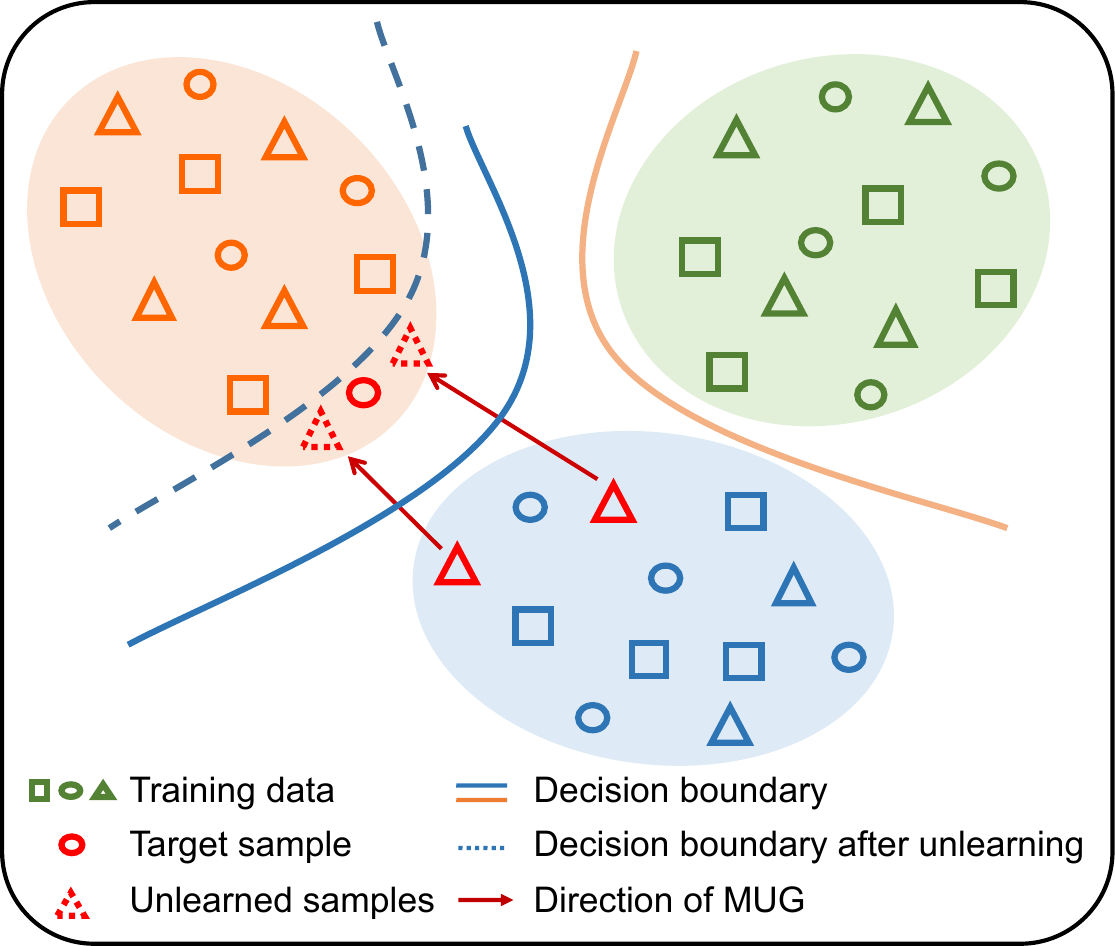}
		\caption{An illustration of FedMUA in the IID setting. The ovals in different colors denote different classes. The solid line represents the decision boundary and the dotted line indicates the decision boundary after unlearning. Moving the unlearned data to the decision boundary for unlearning can substantially alter the decision boundary around the target sample.}

		\label{fig:technique}
	\end{figure}

	The goal of FedMUA is to intentionally manipulate the federated unlearning process, inducing an altered prediction in the resulting unlearned target client model for the target user. To attain this objective, FedMUA comprises the following two fundamental steps (c.f. Fig.~\ref{fig:technique}):
	
	\textbf{Step I. Influential Sample Identification}: We begin by identifying a small subset of training data that can significantly affect the prediction performance on the target data using influence function methods. Subsequently, we carefully select samples with negative IF values and identical labels to bolster the attack capability, thereby ensuring the successful misclassification of the target data.
	
	\textbf{Step II. Malicious Unlearning Generation}: Given these influential data, we proceed to craft malicious unlearning requests by manipulating the features of these influential data to align with those of the target sample. This deliberate manipulation ultimately results in a modification of predictions for the target sample.

	\subsection{Influential Sample Identification}\label{sec-5-1}
	
	To alter the predictions of the target data by the target client's local model, our attack capitalizes on the following intuition: Given a specific target data, it is feasible to manipulate the unlearning process of the federated learning model in a way that hinders it from acquiring useful information from the target data. This manipulation, in turn, leads to different predictions for the target data.
	
	Therefore, it is crucial to identify influential samples within the training dataset that have the most positive impact on the prediction of the target data. To accomplish this, we initiate the process by employing influence function methods to identify these impactful samples. Specifically, we utilize the chain rule to calculate the influence of upweighting a training data $\boldsymbol{z}=(\boldsymbol{x},y)$ from the attacker on the loss at a target data $\boldsymbol{z}_t=(\boldsymbol{x}_t,y)$ as follows:
	\begin{align}
		\label{eqn:influence}
		I_{up,loss}(\boldsymbol{z}, \boldsymbol{z}_t) & = \frac{d \mathcal L(\boldsymbol{z}_t, \boldsymbol{\hat w}_{\epsilon, \boldsymbol{z}})}{d\epsilon}\Bigr|_{\substack{\epsilon = 0}} \\
		& = \nabla_w L(\boldsymbol{z}_t, \boldsymbol{\hat w}) ^\top \frac{d\boldsymbol{\hat w}_{\epsilon, \boldsymbol{z}}}{d\epsilon}\Bigr|_{\substack{\epsilon = 0}} \nonumber \\
		& = -\nabla_w L(\boldsymbol{z}_t, \boldsymbol{\hat w}) ^\top H_{\boldsymbol{\hat w}}^{-1} \nabla_w L(\boldsymbol{z}, \boldsymbol{\hat w}),
		\label{eq:IF}
	\end{align}
	where $\boldsymbol{\hat w}_{\epsilon, \boldsymbol{z}}=\arg\min_{\boldsymbol{w}\in\boldsymbol{\Theta}} \frac{1}{m} \sum_{j=1}^m \mathcal L(\boldsymbol{z}_j,\boldsymbol{w})+ \epsilon \mathcal L(\boldsymbol{z}, \boldsymbol{w})$,
	$H_{\boldsymbol{\hat w}}= \frac{1}{m} \sum_{j=1}^m \nabla^2_w \mathcal L(\boldsymbol{z}_j, \boldsymbol{\hat w})$ is the Hessian, and $\boldsymbol{\hat w} = \arg\min_{\boldsymbol{w}\in\boldsymbol{\Theta}} \frac{1}{m}
	\sum_{j=1}^m \mathcal L(\boldsymbol{z}_j,\boldsymbol{w})$

	Importantly, it is also essential to ensure that the IF value of the loss (denoted as $I_{up,loss}$) associated with the data is negative. The value of $I_{up,loss}$ represents the increase in loss for the target data. Thus, a negative value indicates that the training sample can effectively reduce the loss for the target data, bringing the predicted value of the target data closer to its actual value and thereby enhancing its relevance. Furthermore, a higher negative $I_{up,loss}$ value corresponds to a more substantial impact.
	
	To select these influential samples, we then calculate the value of $I_{up,loss}$ of each sample within the attacker's training dataset $\mathcal{D}_m=\{(\boldsymbol x_j, y_j)\}_{j=1}^m$. Subsequently, we select $n$ samples ($n \leq m$) with the highest negative values of $I_{up,loss}$ in the following manner:
	\begin{align}
		\mathcal{D}_n=\mathop{\max}_{n}(\mathrm{IF}(\mathcal{D}_{m})).
		\label{eq:negative_IF}
	\end{align}

	Considering that each influential sample contributes to the classification of every category, altering these samples corresponding to different categories when attacking the target data will have a counteractive effect. Therefore, to further strengthen our attack, we aim to select $p$ samples ($p \leq n$) with the identical label in $\mathcal{D}_n$ and diminish the influence of samples with different labels identical as follows:
	\begin{align}
		\mathcal{D}_{inf}=\mathrm{S}_p(\mathcal{D}_{n}).
		\label{eq:same_label}
	\end{align}
	where $S(.)$ is the method to find the same label in the training dataset. This step enables the identification of the most influential samples, denoted as $\mathcal{D}_{inf}=\{(\boldsymbol x_j, y_j)\}_{j=1}^p$, for a given target data.

	\begin{algorithm}[t]
		\caption{FedMUA Algorithm}
		\label{alg:algorithm}
		\textbf{Input}: Malicious client $\mathcal{M}_m(.,\textbf{w}_m)$ with its updated algorithm $\mathcal{L}_m$ and training data $\mathcal{D}_m$. Target data $\boldsymbol x_{t}$.

		\begin{algorithmic}[1] 
			\STATE Calculate IF values for each sample in $\mathcal{D}_{m}\leftarrow	Equ.~\ref{eq:IF}$;
			\STATE Select $n$ negative samples: $\mathcal{D}_n \leftarrow Equ.~\ref{eq:negative_IF}$;
			\STATE Select $p$ samples: $\mathcal{D}_{inf}  \leftarrow   Equ.~\ref{eq:same_label}  $ ;
			\FOR{$j=1;j<=p; j++$}
			\STATE  update unlearned sample: $\boldsymbol{x}'_j  \leftarrow Equ.~\ref{eq:x_j}$
			
			\STATE 	satisfy: $\boldsymbol{x}'_j  \leftarrow Equ.~\ref{eq:bound},$ 
			\STATE  satisfy:$ \lVert\boldsymbol \delta_j \rVert \leq \epsilon$
			
			\ENDFOR
		\STATE	Malicious local model unlearn $\{\boldsymbol \delta_j\}_{j=1}^p$: $\mathcal{M}_m(.,\textbf{w}^{'}_m)$ 
			\STATE \textbf{Output}: Updated malicious local model: $\mathcal{M}_m(.,\textbf{w}^{'}_m)$ 
		\end{algorithmic}
	\end{algorithm}
	
	\subsection{Malicious Unlearning Generation}\label{sec-5-2}

	Building upon the identification of $\mathcal{D}_m$, our subsequent step involves formulating a malicious unlearning request for these samples. The most straightforward approach would be to alter the labels of these influential data. However, such direct label modifications could be easily detected by the server's malicious unlearning request detection mechanisms. To address the need for stealthiness in the unlearning request, we propose a strategy wherein the malicious client manipulates the features of the influential samples.
	
	Typically, manipulating the decision boundary around the target sample can lead to corresponding alterations in the predictions made by FL models. Building on this premise, we propose that the malicious user can amplify the unlearning effect of influential samples by intentionally shifting them towards the target sample. In essence, the malicious feature unlearning process is strategically designed to move the influential samples closer to the target sample. Specifically, when the malicious feature unlearning is triggered, the features of these influential samples are adjusted in such a way that they converge towards the features of the target sample. In other words, the samples in proximity to the target sample become these unlearned influential samples, now assigned a different label than that of the target sample. As a result, the prediction for the target sample is inevitably altered, aligning with the attacker's objectives.

	Based on our method, the malicious client endeavors to construct an unlearned version of each sample in $\mathcal{D}_{inf}$:
	\begin{equation}
		\boldsymbol{x}'_j=\boldsymbol{x}_j-\boldsymbol{\delta}_j,
		\label{eq:x_j}
	\end{equation}
where $\boldsymbol{\delta}_j$ is the perturbation.

	Recall that the attacker has two goals. To achieve Goal I, each sample $\boldsymbol{x}^\prime$ in $\mathcal{D}_{inf}$ is required to satisfy:
	\begin{equation}
		\lVert \boldsymbol{x}'_j-\boldsymbol{x}_t \rVert \leq \zeta,
		\label{eq:bound}
	\end{equation}
	where $\zeta$ is a small value of the distance between the unlearned sample and the target data. To achieve Goal II, it is also necessary to satisfy that $\lVert\boldsymbol \delta_j \rVert \leq \epsilon$.

	It is important to note that different values of $\epsilon$ result in varying levels of attack effectiveness. Specifically, a larger $\epsilon$ leads to a smaller $\zeta$, which enhances the attack's strength. However, this increased attack effectiveness comes at the cost of reduced model utility. This trade-off between attack strength and model utility is crucial to consider when adjusting these parameters. We have also evaluated the performance of FedMUA under different values of $\epsilon$ in our experiments.
	After generating and initiating the malicious unlearning data, we can obtain the modified model capable of altering the prediction output for the target data. The entire FedMUA process is encapsulated in Algorithm~\ref{alg:algorithm}.

	\section{Defensive Mechanism}
	\label{sec:DEFENSIVE MECHANISM}
	
In the above section, we introduced a malicious unlearning attacking strategy to demonstrate the vulnerabilities of federated learning during the unlearning process. However, existing defenses are generally designed for traditional attacks~\cite{zhang2024flpurifier,yao2024reverse,jin2022can,li2021anti,chen2021pois}, or are computationally prohibitive for standard machine unlearning pipelines. To address this, we propose a defensive mechanism specifically designed to increase robustness against malicious unlearning attacks.

\begin{figure}[t]
	\centering
	\includegraphics[width=70mm]{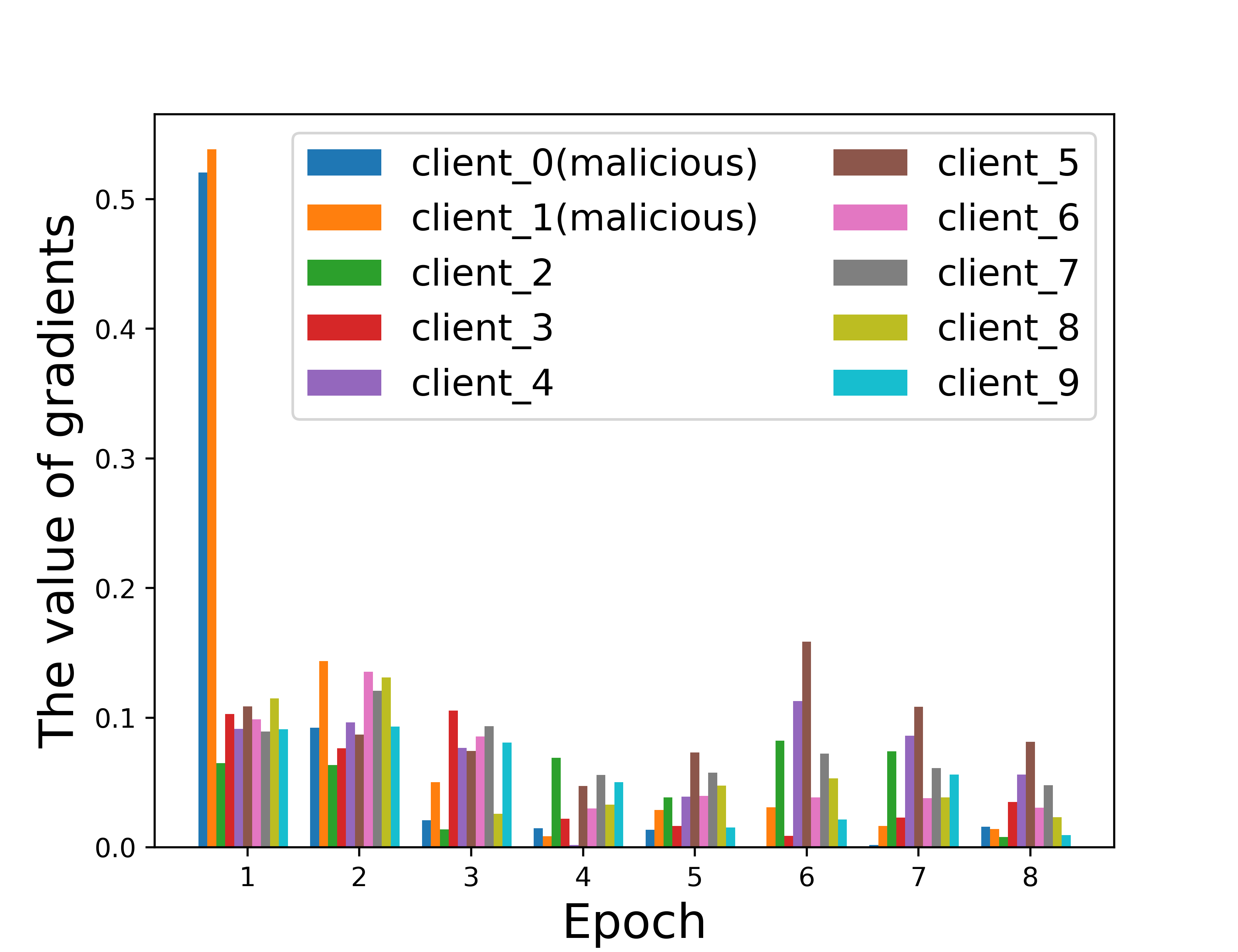}\hspace{2mm}
	\vspace{-5pt}
	\caption{Key observation in MNIST. The value of gradients refers to the sum of the $\ell_2$-norms of each gradient for each client in each training round. Specifically, client 0 and client 1 send malicious unlearning requests to the server. By visualizing the gradient updates for each client in each training round, we observe that the gradient updates from client 0 and client 1 are generally larger than those from normal clients in the initial training rounds.}
	\label{fig:observation}
\end{figure}

Specifically, as illustrated in Fig.~\ref{fig:observation}, we visualize the gradient updates for each client in each training round using the MNIST dataset, we can observe that the gradient updates from malicious clients (i.e., client 0 and client 1) are generally larger than those of normal clients in the initial training rounds. Based on this observation, we propose a novel defense method, a rapid yet effective way to identify malicious unlearning requests and reduce their negative impact on the global model for a given client. 

To achieve this goal, our defense method mainly consists of two steps. First, we identify malicious unlearning requests that can significantly affect the prediction performance on the target data. These requests can be more rapidly found out based on their larger gradient updates to the server. In this step, we use the Interquartile Range (IQR) method~\cite{aggarwal2017introduction}, a common technique to identify gradient values that are abnormally distant from most of the other gradients. Those gradient updates that are larger than the upper quartile can be regarded as malicious unlearning requests. Once these malicious unlearning requests are identified, the next step is to reduce the value of these gradients by multiplying them by a fixed parameter $\lambda$ (e.g., 0.1 or 0.5 in our experiments) to minimize their negative impact on the global model. In this way, we can finally mitigate malicious unlearning attacks on the target sample.

	\section{Evaluation}
	\label{sec:Evaluation}
	
	In this section, we first provide details of the datasets, evaluation metrics, and aggregation rules used in our experiments. Then, we conduct experiments in both IID and Non-IID settings to evaluate the performance of FedMUA.

	\subsection{Experiment Setup} \label{Experimental Setup}

	\subsubsection{Datasets} \label{Dataset Description}
	We evaluate FedMUA using five datasets in our experiments including: Purchase\footnote{\url{https://www.kaggle.com/c/acquire-valued-shoppers-challenge/data}}, MNIST\footnote{\url{http://yann.lecun.com/exdb/mnist}}, CIFAR-10\footnote{\url{http://www.cs.toronto.edu/~kriz/cifar.html}},
	CIFAR-100\footnote{\url{http://www.cs.toronto.edu/~kriz/cifar.html}}, and
	Credit Score\footnote{\url{https://www.kaggle.com/datasets/parisrohan/credit-score-classification}}.
	
	\textbf{\emph{Purchase.}}  Purchase dataset is sourced from Kaggle's "Acquire Valued Shoppers" challenge, aiming to develop precise coupon promotion strategies. Specifically, it consists of $197,324$ records, lacks predefined class labels. Consistent with~\cite{shokri2017membership,salem2019ml}, we employ an unsupervised clustering algorithm to assign class labels to each data record. The dataset is clustered into 2 classes.

	\textbf{\emph{MNIST.}} It comprises $70,000$ handwritten digits presented as $28 \times 28$ images. It has been normalized, placing the digits at the center of the image. The dataset encompasses samples of handwritten digits spanning from $0$ to $9$, with each pixel represented by a binary value (0 or 1).

	\textbf{\emph{CIFAR-10.}} It is a widely utilized benchmark dataset for assessing image recognition algorithms. It comprises $60,000$ color images sized $32\times32$, categorized into $10$ classes. Notably, this dataset maintains balance, with $6,000$ randomly chosen images per class. It is divided into $50,000$ training images and $10,000$ testing images.
	
    \textbf{\emph{CIFAR-100.}} It consists of 20 major categories and a total of 100 subcategories. For each subcategory, there are 500 training samples and 100 testing samples, with each sample having dimensions of $32\times32$.
	
     \textbf{\emph{Credit Score.}} It aims to build a machine learning model to classify people into different credit scores. Specifically, the original data consists of 50,000 real credit records and each record contains a total of 27 features. After data preprocessing, the total data expands to 97,799 samples, with each sample containing 34 features.

	\subsubsection{Evaluation Metrics} \label{Metrics}
	We adopt the following three metrics to evaluate the performance of FedMUA: 
	
	\textbf{Attack Success Rate (ASR)}. Given the target testing samples, ASR represents the ratio of misclassified samples by the unlearned target client model $\widetilde{\mathcal{M}_t}$. The formulation is as follows:
	\begin{align}
		ASR = \frac{\sum_{(\mathbf{x}_t,y_t)\in\mathcal{D}_t}\mathbb{I}(\widetilde{\mathcal{M}_t}( \mathbf{x}_t) \neq y_t)}{m},
	\end{align}
	where $\mathbb{I}$ is an indicator function, and $\mathcal{D}_t=\{(\mathbf{x}_t,y_t)\}_{t=1}^m$ is the set of target samples from the testing dataset. ASR-B represents the ASR that is caused directly by the process of federated unlearning.
	
	\textbf{Unlearned Global Test Accuracy ($\widetilde{Acc}_G$):} $\widetilde{Acc}_G$ is the fraction of testing examples that are correctly predicted by the unlearned global model $\widetilde{\mathcal{M}_G}$, and can be expressed as:
	\begin{align}
			\widetilde{Acc}_G= \frac{\sum_{(\mathbf{x}_i,y_i)\in \mathcal{D}_{test}}\mathbb{I}(\widetilde{\mathcal{M}_G}(\mathbf{x}_i)=y_i)}{n}.
		\end{align}
		where $\mathcal{D}_{test}=\{(\mathbf{x}_i,y_i)\}_{i=1}^n$ is the testing dataset.
		
		\textbf{Clean Global Test Accuracy ($Acc_G$)}. $Acc_G$ is the fraction of testing examples that are correctly predicted by the global model $\mathcal{M}_G$, which is given by:
		\begin{align}
				Acc_G= \frac{\sum_{(\mathbf{x}_i,y_i)\in \mathcal{D}_{test}}\mathbb{I}( \mathcal{M}_G (\mathbf{x}_i)=y_i)}{n}.
			\end{align}

			\begin{table*} [t]
	\centering 
	\caption{Performance of FedMUA under various FU methods and aggregation rules across three datasets in the IID setting.}
	\resizebox{1.01\textwidth}{!}{
		{         
			\begin{tabular}{@{}c|c|cccc|cccc|cccc@{}}
				\toprule
				\multirow{2}{*}{\textbf{FU}} & \multirow{2}{*}{\textbf{Aggregation}} & \multicolumn{4}{c|}{\textbf{Purchase}}& \multicolumn{4}{c|}{\textbf{MNIST}}& \multicolumn{4}{c}{\textbf{CIFAR-10}}  \\ \cmidrule(l){3-14} 
				\textbf{Method}	&\textbf{Rule} & ASR-B & ASR\phantom{$\uparrow$}&$Acc_G$(\%)&$\widetilde{Acc}_G$ (\%)& ASR-B & ASR &$Acc_G$(\%)\phantom{$\uparrow$}& $\widetilde{Acc}_G$(\%)\phantom{$\uparrow$}& ASR-B & ASR\phantom{$\uparrow$} &$Acc_G$(\%)& $\widetilde{Acc}_G$ (\%) \\ \midrule
				\multirow{3}{*}{FedEraser~\cite{liu2021federaser}} 
				& FedAvg& 0.05& 0.90 \phantom{$\uparrow$}&94.09 $\pm$ 0.06  & 94.23 $\pm$ 0.19 & 0.00& 0.85&97.39 $\pm$ 0.06 $\uparrow$  & 97.27 $\pm$ 0.07 \phantom{$\uparrow$}& 0.00& 0.95 $\uparrow$&84.20 $\pm$ 0.40  & 83.30 $\pm$ 0.26\\
				& Median& 0.05& 0.90 $\uparrow$&94.44 $\pm$ 0.03  &94.51 $\pm$ 0.21& 0.00&0.75&97.40 $\pm$ 0.06 $\uparrow$& 97.19 $\pm$ 0.10 \phantom{$\uparrow$}& 0.00&0.85 \phantom{$\uparrow$}&83.57 $\pm$ 0.55 & 81.43 $\pm$ 0.40 \\
				& Trimmed-mean & 0.05& 0.90 \phantom{$\uparrow$}&94.30 $\pm$ 0.04   & 94.33 $\pm$ 0.22 & 0.00& 0.80& 97.39 $\pm$ 0.03 $\uparrow$&  97.27 $\pm$ 0.09 \phantom{$\uparrow$}& 0.05& 0.95 $\uparrow$& 84.25 $\pm$ 0.24& 83.32 $\pm$ 0.26  \\ 
				& Krum & 0.05& 0.85 \phantom{$\uparrow$}&94.10 $\pm$ 0.14   & 94.21 $\pm$ 0.26 & 0.00& 0.80& 97.37 $\pm$ 0.04 $\uparrow$&  97.21 $\pm$ 0.09 \phantom{$\uparrow$}& 0.05& 0.90 $\uparrow$& 84.02 $\pm$ 0.28& 83.15 $\pm$ 0.28 \\ \midrule 
				
				\multirow{3}{*}{KNOT~\cite{su2023asynchronous} } 
				& FedAvg  & 0.15& 0.95 $\uparrow$&81.22 $\pm$ 0.03  & 81.03 $\pm$ 0.01& 0.00& 0.90&  97.31 $\pm$ 0.02 $\uparrow$& 96.79 $\pm$ 0.05 \phantom{$\uparrow$}& 0.10& 0.95 $\uparrow$& 80.51 $\pm$ 0.04& 80.61 $\pm$ 0.07\\
				& Median & 0.10& 0.95 $\uparrow$&81.09 $\pm$ 0.04 & 81.39 $\pm$ 0.20 & 0.00& 0.90& 97.11 $\pm$ 0.05 $\uparrow$&97.11 $\pm$ 0.02 \phantom{$\uparrow$}& 0.05& 0.90 \phantom{$\uparrow$}& 80.49 $\pm$ 0.06& 80.60 $\pm$ 0.06 \\
				& Trimmed-mean & 0.10& 0.95 $\uparrow$&81.50 $\pm$ 0.04  & 81.04 $\pm$ 0.09 & 0.05& 0.80 & 97.08 $\pm$ 0.01 \phantom{$\uparrow$}&97.14 $\pm$ 0.01 $\uparrow$& 0.05&0.95 $\uparrow$& 80.43 $\pm$ 0.03 & 80.57 $\pm$ 0.06\\ 
				& Krum & 0.05& 0.90 \phantom{$\uparrow$}&80.99 $\pm$ 0.24  & 81.05 $\pm$ 0.21 & 0.00& 0.90 & 96.40 $\pm$ 0.07 $\uparrow$&96.31 $\pm$ 0.01 \phantom{$\uparrow$}& 0.05&0.95 $\uparrow$& 80.67 $\pm$ 0.08 & 79.07 $\pm$ 0.08\\ \midrule
				
				\multicolumn{2}{c|}{Average}  & 0.075&0.912 \phantom{$\uparrow$}  &87.71 $\pm$ 0.08  & 87.72 $\pm$ 0.17 & 0.006&0.837 &97.18$\pm$ 0.04 \phantom{$\uparrow$} & 97.03 $\pm$ 0.05 \phantom{$\uparrow$}& 0.044& 0.925\phantom{$\uparrow$} & 82.29 $\pm$ 0.21 & 81.50 $\pm$ 0.18  \\ \bottomrule 
			\end{tabular}
			\label{tab:iid}
		}
	}
\end{table*}

			\begin{table*} [t]
				\centering 
				\caption{Performance of FedMUA under various FU methods and aggregation rules across three datasets in the Non-IID setting.}
				\resizebox{1.01\textwidth}{!}{
					{         
						\begin{tabular}{@{}c|c|cccc|cccc|cccc@{}}
							\toprule
							\multirow{2}{*}{\textbf{FU}} & \multirow{2}{*}{\textbf{Aggregation}} & \multicolumn{4}{c|}{\textbf{Purchase}}& \multicolumn{4}{c|}{\textbf{MNIST}}& \multicolumn{4}{c}{\textbf{CIFAR-10}}  \\ \cmidrule(l){3-14} 
							\textbf{Method}	&\textbf{Rule} & ASR-B & ASR &$Acc_G$(\%)&$\widetilde{Acc}_G$ (\%)& ASR-B& ASR &$Acc_G$(\%) \phantom{$\uparrow$}& $\widetilde{Acc}_G$(\%) \phantom{$\uparrow$}& ASR-B& ASR &$Acc_G$(\%)& $\widetilde{Acc}_G$ (\%) \\ \midrule
							\multirow{3}{*}{FedEraser~\cite{liu2021federaser}} 
							& FedAvg& 0.05& 0.85 \phantom{$\uparrow$}&93.67 $\pm$ 0.18  & 93.52 $\pm$ 0.27 & 0.00& 0.45& 96.23 $\pm$ 0.03  \phantom{$\uparrow$}& 96.75 $\pm$ 0.09 $\uparrow$& 0.05& 0.90 $\uparrow$&82.35 $\pm$ 0.25  & 82.05 $\pm$ 0.21\\
							& Median& 0.05& 0.85 $\uparrow$&94.04 $\pm$ 0.23  &93.69 $\pm$ 0.25&0.00&0.50&95.59 $\pm$ 0.08 \phantom{$\uparrow$}& 96.39 $\pm$ 0.09 $\uparrow$ & 0.00& 0.80 \phantom{$\uparrow$}&80.45 $\pm$ 0.40 & 80.58 $\pm$ 0.42 \\
							& Trimmed-mean & 0.00& 0.80 \phantom{$\uparrow$}&93.88 $\pm$ 0.24   & 93.61 $\pm$ 0.33 & 0.00& 0.50& 96.26 $\pm$ 0.03 \phantom{$\uparrow$} &  96.71 $\pm$ 0.07 $\uparrow$ & 0.05& 0.85 $\uparrow$& 82.24 $\pm$ 0.42& 82.19 $\pm$ 0.37 \\ 
							& Krum & 0.05& 0.80 \phantom{$\uparrow$}&93.41 $\pm$ 0.23  & 93.18 $\pm$ 0.23& 0.00& 0.45&95.94 $\pm$ 0.09 $\uparrow$&95.08 $\pm$ 0.12 \phantom{$\uparrow$}& 0.05&0.85 $\uparrow$&82.08 $\pm$ 0.23& 81.53 $\pm$ 0.26 \\
							\midrule 
							
							\multirow{3}{*}{KNOT~\cite{su2023asynchronous} } 
							& FedAvg & 0.00 & 0.55  \phantom{$\uparrow$} &60.99 $\pm$ 0.05  & 60.32 $\pm$ 0.07& 0.00& 0.55&96.74 $\pm$ 0.42 $\uparrow$& 95.93 $\pm$ 0.05 \phantom{$\uparrow$}& 0.10& 0.90 $\uparrow$&  63.49 $\pm$ 0.05 & 65.32 $\pm$ 0.07\\
							& Median & 0.00& 0.60  \phantom{$\uparrow$}&60.13 $\pm$ 0.01  & 60.31 $\pm$ 0.15& 0.00& 0.55 & 95.97 $\pm$ 0.02 $\uparrow$&95.76 $\pm$ 0.11 \phantom{$\uparrow$}& 0.00&0.75 $\uparrow$& 65.60 $\pm$ 0.73& 65.44 $\pm$ 0.48   \\
							& Trimmed-mean & 0.00& 0.60 \phantom{$\uparrow$} &61.01 $\pm$ 0.21  & 60.68 $\pm$ 0.07& 0.00& 0.45&95.86 $\pm$ 0.02 \phantom{$\uparrow$}&97.14 $\pm$ 0.02 $\uparrow$& 0.05&0.85 $\uparrow$&64.29 $\pm$ 0.48& 65.49 $\pm$ 0.45 \\ 
							& Krum & 0.00& 0.55  \phantom{$\uparrow$}&60.42 $\pm$ 0.13  & 60.58 $\pm$ 0.19& 0.00& 0.45&95.74 $\pm$ 0.29 $\uparrow$&95.08 $\pm$ 0.38 \phantom{$\uparrow$}& 0.05&0.80 $\uparrow$&63.08 $\pm$ 0.22& 61.56 $\pm$ 0.16 \\
							\midrule
							
							\multicolumn{2}{c|}{Average}  & 0.018& 0.70  \phantom{$\uparrow$}&77.19 $\pm$ 0.16  &76.98 $\pm$ 0.19  & 0.00& 0.487 &96.04$\pm$ 0.12 \phantom{$\uparrow$}&96.10 $\pm$ 0.12 \phantom{$\uparrow$}  & 0.04& 0.84  \phantom{$\uparrow$}& 72.94 $\pm$ 0.34 & 73.02 $\pm$ 0.30 \\ \bottomrule 
						\end{tabular}
						\label{tab:non-iid}
					}
				}
			\end{table*}
			
			
			\subsubsection{FU Methods and Aggregation Rules} \label{Aggregation Rules}
			
			We first employ two FU methods, namely FedEraser~\cite{liu2021federaser} and KNOT~\cite{su2023asynchronous}, both of which have been extensively discussed in the related work. The selection of these methods is motivated by their prominence in the field, aligning with the objectives of our study. Then, we will discuss four commonly used aggregation rules, including the non-robust FedAvg~\cite{mcmahan2017communication}, and three Byzantine-robust ones, i.e., Median~\cite{yin2018byzantine}, Trimmed-mean~\cite{yin2018byzantine}  and Krum~\cite{blanchard2017machine}. 
			
			\textbf{FedAvg}: FedAvg~\cite{mcmahan2017communication} stands out as the most widely adopted aggregation rule in FL. This method computes the global model update by averaging the local model updates. FedAvg has demonstrated state-of-the-art performance in non-adversarial scenarios.

			\textbf{Median}: Median~\cite{yin2018byzantine} represents a coordinate-wise aggregation approach in FL. In this method, the server sorts the parameter values in local model updates and identifies the median value for each parameter, which then serves as the aggregated value in the global model update.

			\textbf{Trimmed-mean}: Trimmed-mean~\cite{yin2018byzantine} involves removing the largest and smallest k values from the sorted parameter values. Subsequently, it calculates the average of the remaining values, representing the corresponding parameter in the global model update. The parameter kis a parameter that balances the removal of extreme values and the preservation of valuable information.
			
		\textbf{Krum}: Krum~\cite{blanchard2017machine} aims to choose a local model that is similar to the majority of $n$ local models, thereby minimizing the influence of any potentially malicious models. Specifically, $n-m-2$ local models are selected to calculate their Euclidean distance with local model and Krum selects one with the smallest sum of distance as the global model. Here, $m$ is the number of malicious clients.

			\subsubsection{Baseline} \label{Baseline}
		 Given the absence of prior research on studying the vulnerabilities of federated learning to malicious unlearning attacks, we employ the following three baselines: 
		
		\textbf{Rand+MUG}: This baseline is formed by an intuitive method that incorporates attacks using a random selection of instances for unlearning. Recall that FedMUA comprises two components: \textbf{ISI+MUG}. For comparison, we randomly select influential samples (referred to as `\textbf{Rand}') to replace ISI.
		
		\textbf{HP~\cite{di2022hidden}}: HP was initially designed for typical machine learning. Specifically, they first inject both poisoned and camouflage sets into the training dataset. An unlearning request is then triggered to remove the camouflage dataset, thereby activating the poison effect. We extend HP to the federated learning setting and utilize their method to craft poisoned and camouflage sets for malicious clients.
		
		\textbf{MSFA~\cite{zhao2024static}}: MSFA was originally designed for typical machine learning and they formulate the attack as a bi-level optimization. Specifically, the unlearning samples $D_{f}$ can be optimized by  $D_{f}=\arg \min _{D_{f} \subset D_{t}} \mathcal{L}_{a d v}\left(\cdot ; \theta^{u}\left(D_{f}\right)\right)$ and $\theta^{u}\left(D_{f}\right)=U_{A}\left(D, f_{D}\left(\theta^{*}\right), D_{f}=D_{t} \cap \Omega\right)$, where 
		$D_{t}=\left\{\left(x_{p}, y_{p}\right)\right\}_{p=1}^{P}, \Omega=\left\{\omega_{p} \in \left\{0,1 \right\}\right\}_{p=1}^{P}$. $U_{A}$ is the unlearning algorithm, $\theta^{u}\left(D_{f}\right)$ is the model parameter, $\mathcal{L}_{a d v}$ is the loss function, $f_{D}\left(\theta^{*}\right)$ is the model and $D$ is the training dataset. We thus extend MSFA to the federated learning setting. In this setting, $f_{D}\left(\theta^{*}\right)$ is the global model and $D$ is the dataset for the malicious client.

   To the best of my knowledge, there are currently no defense methods against malicious unlearning attacks. In this context, we explore two defenses beyond gradient monitoring and use them as baseline defenses.
	
	\textbf{FAT~\cite{zizzo2020fat}}: Adversarial training is a common approach to enhancing a model's robustness against adversarial attacks. FAT examined the effectiveness of the federated adversarial training protocol within a FL setting and demonstrated the performance of their models in a distributed FL environment. Since the malicious unlearning requests generated by FedMUA can be considered as adversarial data, FAT can be then employed as a potential defense against FedMUA.
	
	\textbf{FADngs~\cite{dong2024fadngs}}: Anomaly detection seeks to identify anomalous behavior. FADngs trains local models using a contrastive learning approach, which develops more discriminative representations tailored for anomaly detection by utilizing shared density functions.
		
			\subsubsection{Parameter Setting} \label{Experimental Setting}
			Our methods and baseline are implemented in Python 3.8 and utilize the PyTorch library.
			All experiments are performed on a workstation equipped with one NVIDIA GeForce RTX 4090 GPU. We adopt the following parameter settings for the original FL training: In FedEraser, we use 3 FC layers model for purchase dataset, the 2 Conv. and 2 FC layers model for MNIST dataset, and the VGG~\cite{simonyan2015very} model for CIFAR-10 dataset. We train model from scratch for 20 epochs using SGD with a fixed learning rate of 0.005, momentum of 0.9 and batch-size of 64. We set the number of clients to 20 and 10 clients' training data are used in each training round. In KNOT, we utilize a 5 FC layers model for purchase dataset, the LeNet5 model for MNIST dataset, and the ResNet18~\cite{he2016deep} model for CIFAR-10 dataset. We train the model from scratch for 10 local epochs and 20 global epochs with a fixed learning rate of 0.01, momentum of 0.9 and batch-size of 64. We set the number of clients to 10 in each dataset. In the non-IID setting, we use a typical data partitioning strategy called Dirichlet sampled data, each local client is assigned a portion of the samples for each label based on a Dirichlet distribution~\cite{hsu2019measuring} with a concentration parameter of 0.5. We randomly select 40 targets for experiments and report the average results.

			\begin{figure*}[t]
				\centering
				\subfigure[Purchase]
				{\includegraphics[width=56mm]{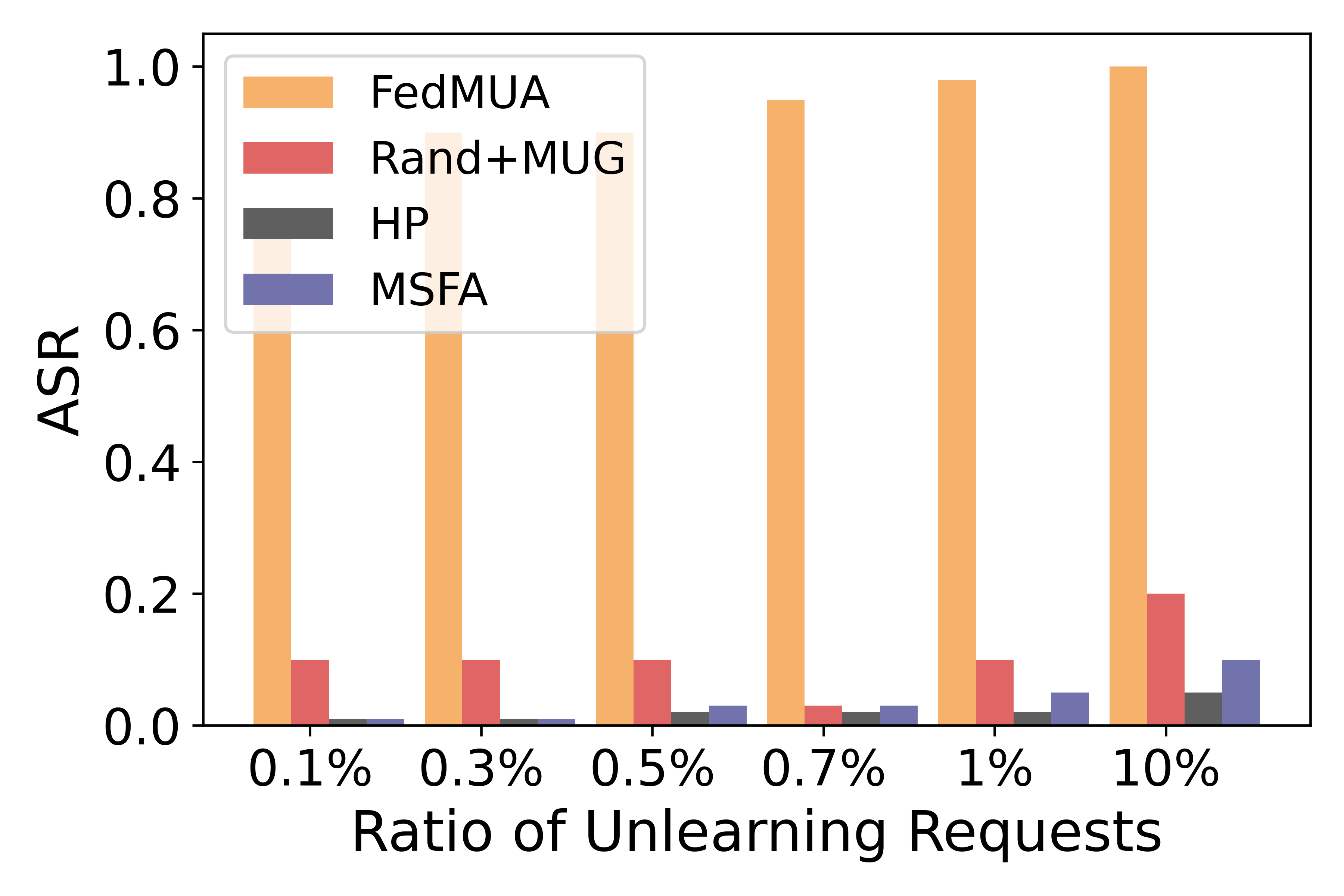}}\label{fig:a}\hspace{2mm}
				\subfigure[MNIST]
				{\includegraphics[width=56mm]{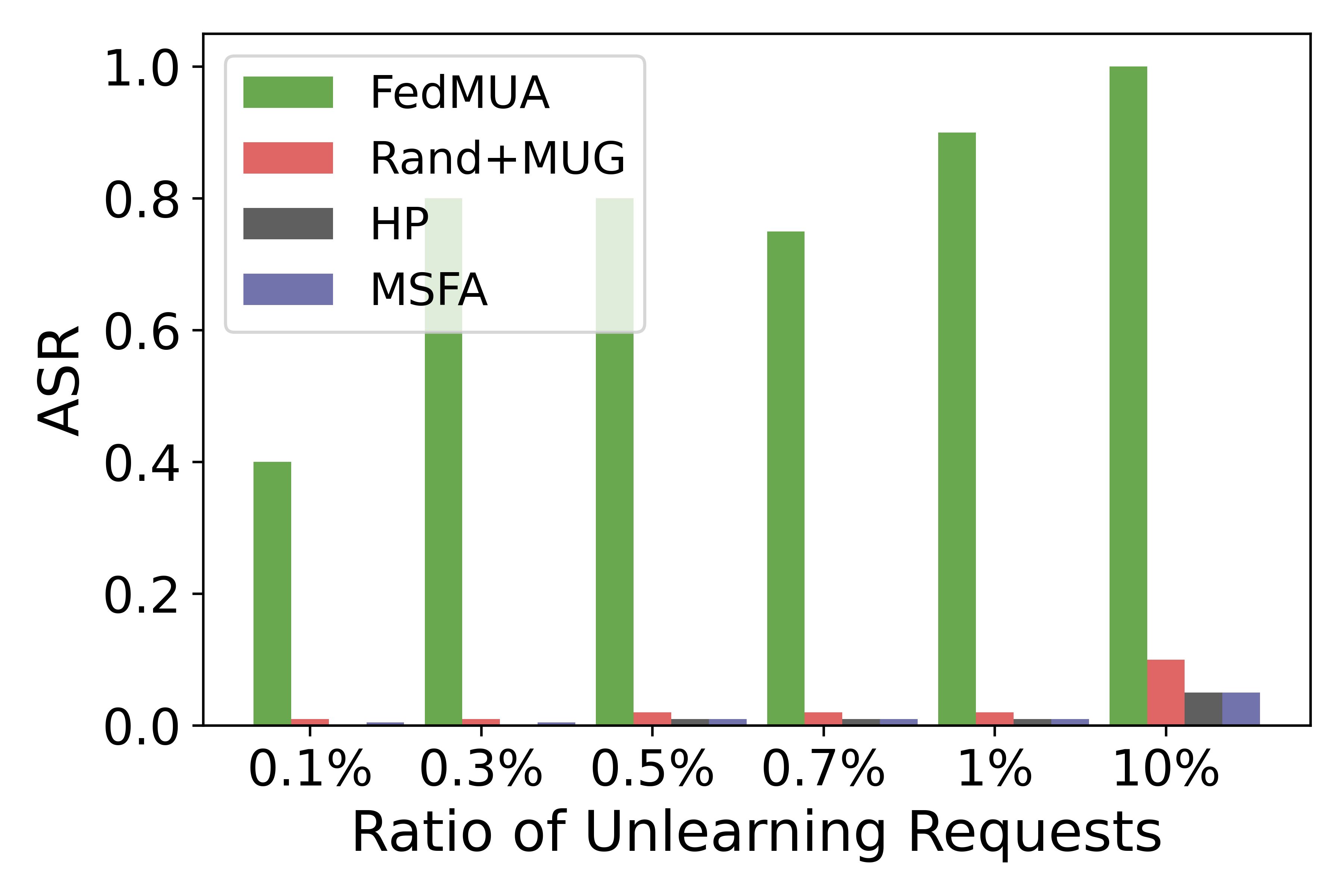}}\label{fig:b}\hspace{2mm}
				\subfigure[CIFAR-10]
				{\includegraphics[width=56mm]{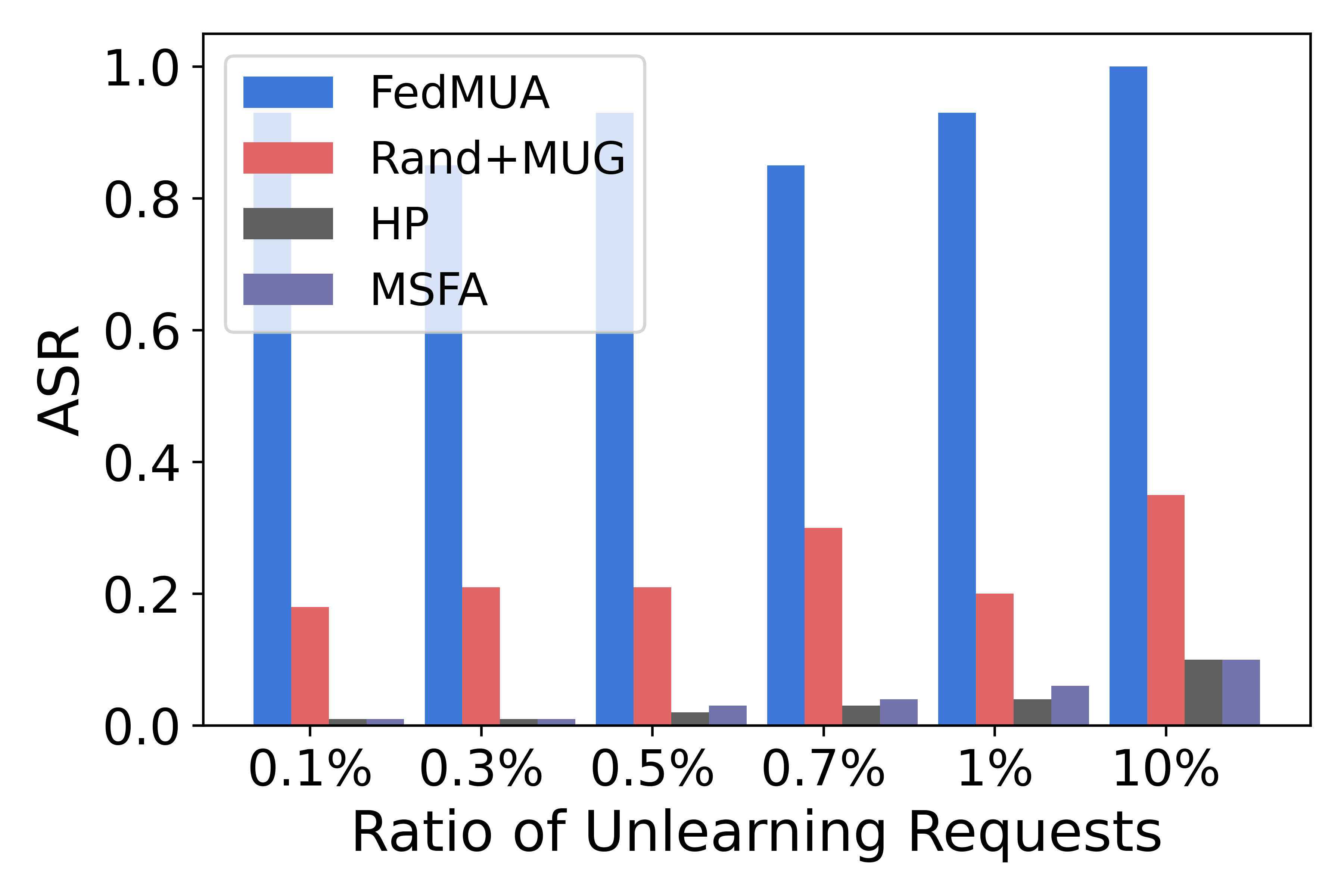}}\label{fig:c}
				\vspace{-5pt}
				\caption{ASR of FedMUA and baseline on FedEraser.}
				\label{fig:ASR_baseline_FedEraser}
				\subfigure[Purchase]
				{\includegraphics[width=56mm]{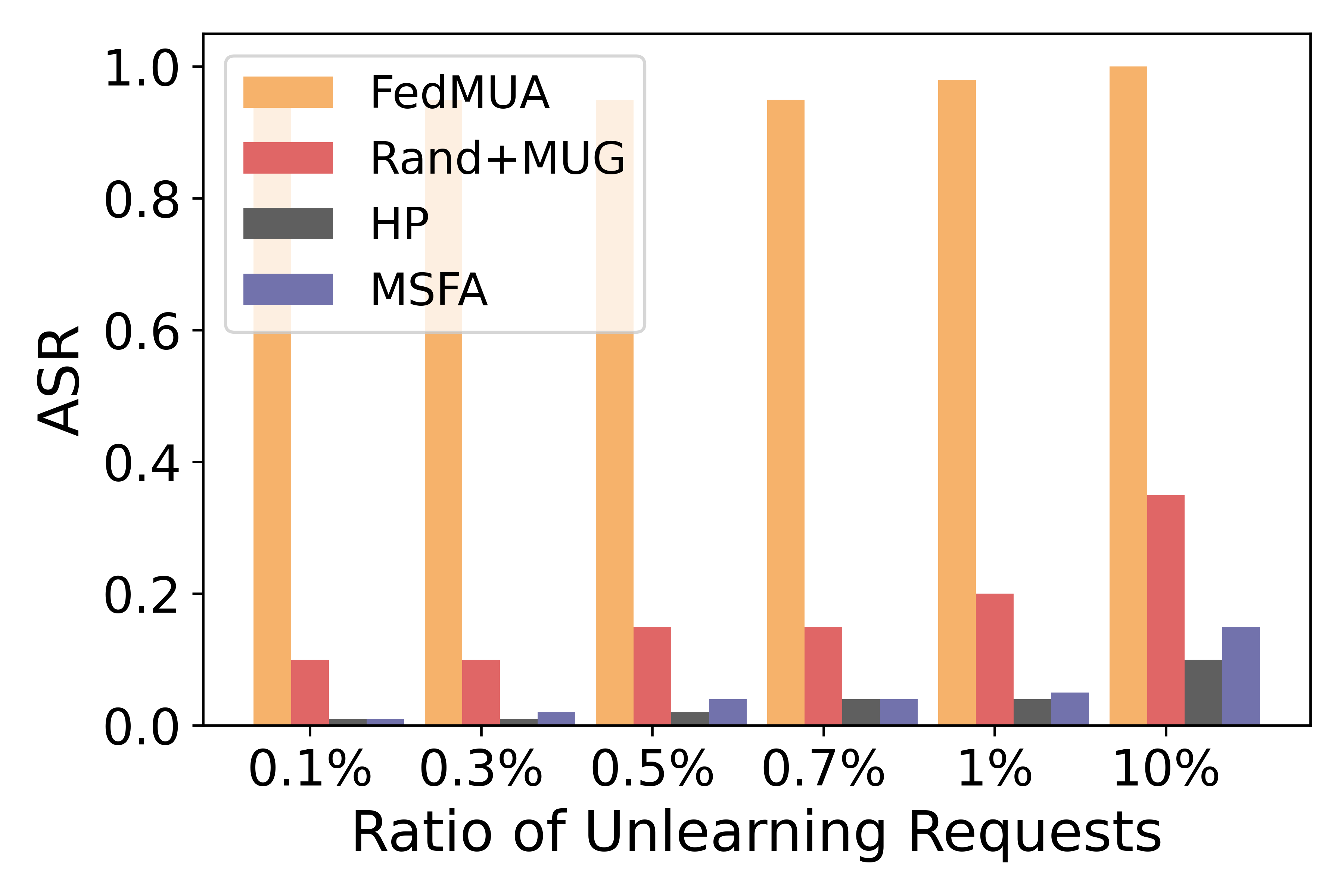}}\label{fig:a}\hspace{2mm}
				\subfigure[MNIST]
				{\includegraphics[width=56mm]{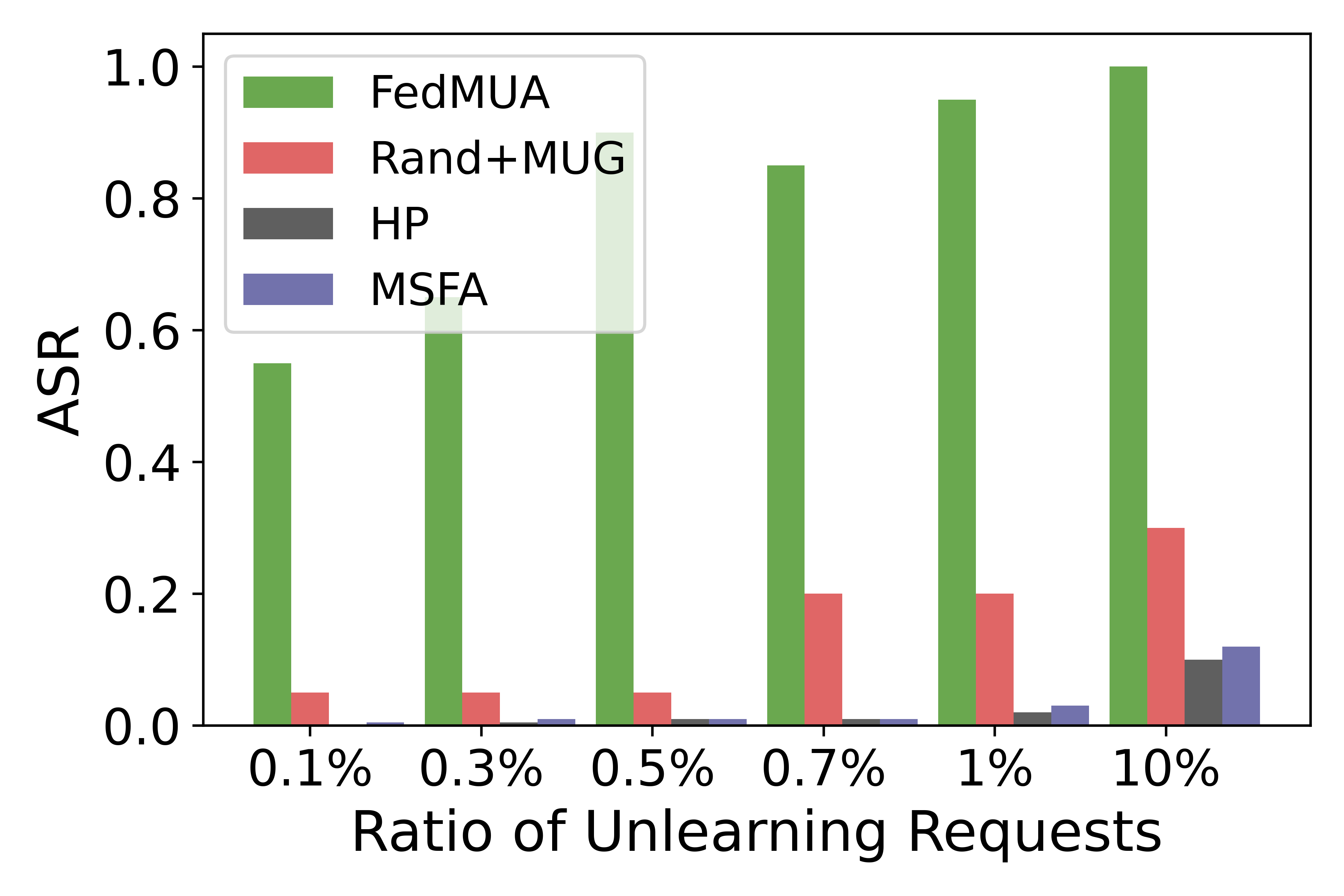}}\label{fig:b}\hspace{2mm}
				\subfigure[CIFAR-10]
				{\includegraphics[width=56mm]{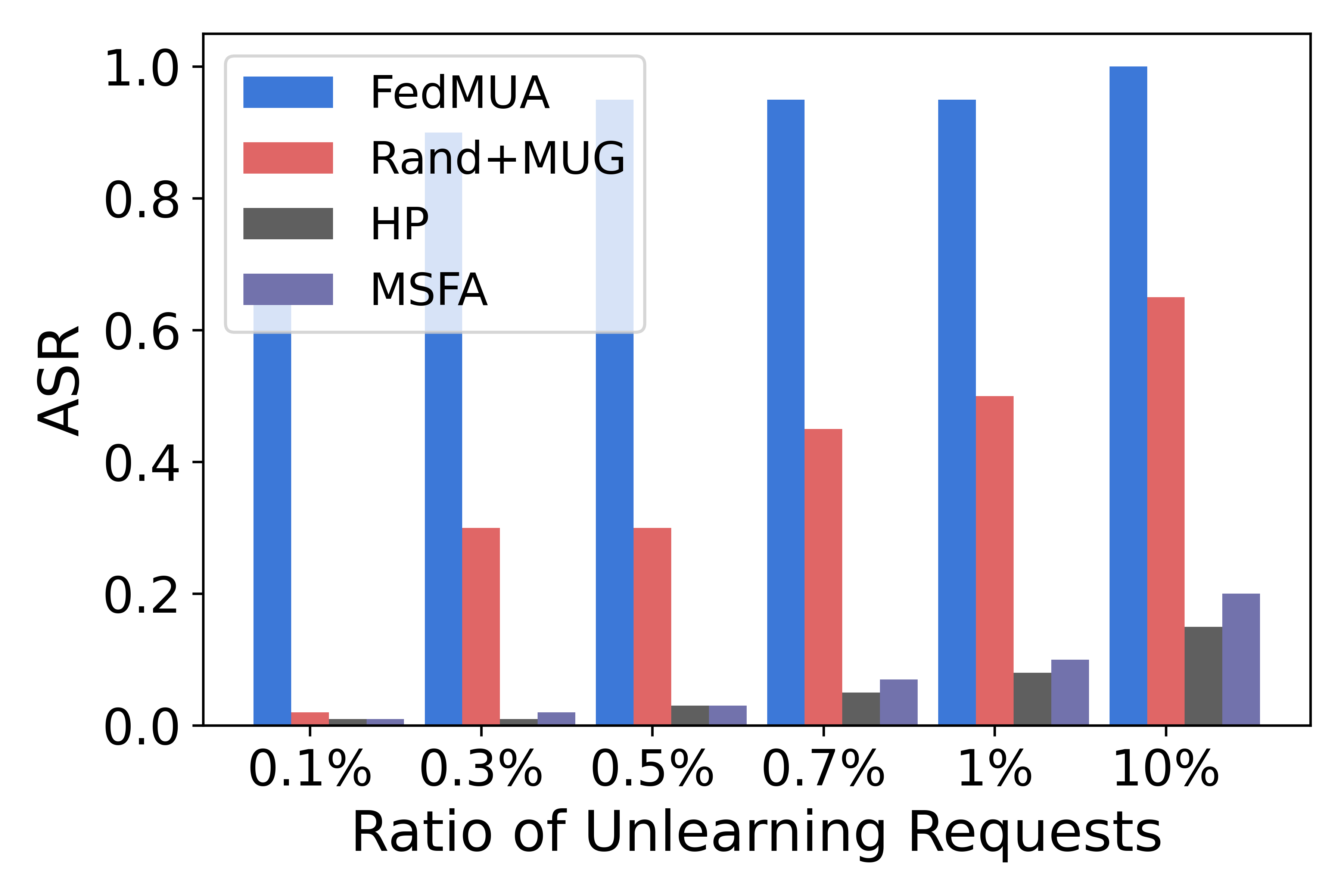}}\label{fig:c}
				\vspace{-5pt}
				\caption{ASR of FedMUA and baseline on KNOT.}
				\label{fig:ASR_baseline}
			\end{figure*}

\begin{table*} [t]
	\centering 
	\caption{Performance comparison between FedMUA and the baselines.}
	\resizebox{1.01\textwidth}{!}{
		{         
			\begin{tabular}{@{}c|c|ccc|ccc|ccc@{}}
				\toprule
				\multirow{2}{*}{\textbf{FU}} & \multirow{2}{*}{\textbf{Ratio of}} & \multicolumn{3}{c|}{\textbf{Purchase}}& \multicolumn{3}{c|}{\textbf{MNIST}}& \multicolumn{3}{c}{\textbf{CIFAR-10}}  \\ \cmidrule(l){3-11} 
				\textbf{Method}	&\textbf{Unlearning Requests} &$Acc_G$(\%)&$\widetilde{Acc}_{G-B}$ (\%) &$\widetilde{Acc}_G$ (\%) &$Acc_G$(\%)\phantom{$\uparrow$}&$\widetilde{Acc}_{G-B}$ (\%)& $\widetilde{Acc}_G$(\%)&$Acc_G$(\%)&$\widetilde{Acc}_{G-B}$ (\%)& $\widetilde{Acc}_G$ (\%) \\ \midrule
				\multirow{5}{*}{FedEraser~\cite{liu2021federaser}} 
				& 0.1\%&94.23 $\pm$ 0.18  & 94.06 $\pm$ 0.15 &94.16 $\pm$ 0.17  & 97.41 $\pm$ 0.13 $\uparrow$&97.11 $\pm$ 0.13  & 97.15 $\pm$ 0.11&83.88$\pm$ 0.21& 83.56 $\pm$ 0.19 &83.67 $\pm$ 0.20\\
				& 0.3\%&94.52 $\pm$ 0.24  &94.23 $\pm$ 0.17&94.48 $\pm$ 0.32& 97.48 $\pm$ 0.15 $\uparrow$&97.30 $\pm$ 0.17 & 97.36 $\pm$ 0.07 &83.75$\pm$ 0.18  & 83.38 $\pm$ 0.22 &83.46 $\pm$ 0.20 \\
				& 0.5\% &94.35 $\pm$ 0.25   & 94.01 $\pm$ 0.21& 94.08 $\pm$ 0.24 &  97.45 $\pm$ 0.15 $\uparrow$& 97.28 $\pm$ 0.15& 97.31 $\pm$ 0.06 &83.50$\pm$ 0.17  & 83.27 $\pm$ 0.20 &83.29 $\pm$ 0.22  \\ 
				& 0.7\%&94.49 $\pm$ 0.22  &94.22 $\pm$ 0.18&94.36 $\pm$ 0.06& 97.50 $\pm$ 0.12 $\uparrow$&97.27 $\pm$ 0.16 & 97.34 $\pm$ 0.13  &83.77$\pm$ 0.25 & 83.43 $\pm$ 0.28 &83.52 $\pm$ 0.39  \\
				& 1\% &94.46 $\pm$ 0.17 & 94.36 $\pm$ 0.22 & 94.44 $\pm$ 0.17 & 97.42 $\pm$ 0.13 $\uparrow$& 97.24 $\pm$ 0.15& 97.29 $\pm$ 0.09 &83.62$\pm$ 0.25  & 83.30 $\pm$ 0.22 &83.37 $\pm$ 0.33  \\ \midrule 
				
				\multirow{5}{*}{KNOT~\cite{su2023asynchronous} } 
				& 0.1\%&81.56 $\pm$ 0.11 & 81.32 $\pm$ 0.10 &81.39 $\pm$ 0.08  & 97.58 $\pm$ 0.05 $\uparrow$&97.27 $\pm$ 0.04& 97.30 $\pm$ 0.02& 80.71 $\pm$ 0.08  & 80.52 $\pm$ 0.08  &80.58 $\pm$ 0.06 \\ 
				& 0.3\%&81.42 $\pm$ 0.15  &81.09 $\pm$ 0.15&81.12 $\pm$ 0.16 & 97.51 $\pm$ 0.05 $\uparrow$&97.23 $\pm$ 0.05 & 97.25 $\pm$ 0.03 &80.72 $\pm$ 0.11  & 80.40 $\pm$ 0.09&80.46 $\pm$ 0.08 \\
				& 0.5\% &81.47 $\pm$ 0.10   & 81.20 $\pm$ 0.10& 81.22 $\pm$ 0.08  &  97.42 $\pm$ 0.06 $\uparrow$& 97.11 $\pm$ 0.04& 97.14 $\pm$ 0.02 &80.65 $\pm$ 0.08  & 80.30 $\pm$ 0.11&80.36 $\pm$ 0.06  \\ 
				& 0.7\%&81.51 $\pm$ 0.09  &81.27 $\pm$ 0.08&81.30 $\pm$ 0.09 & 97.45 $\pm$ 0.06 $\uparrow$&97.10 $\pm$ 0.04 & 97.12 $\pm$ 0.04 &80.74 $\pm$ 0.11  & 80.40 $\pm$ 0.07 &80.43 $\pm$ 0.08 \\
				& 1\% &81.49 $\pm$ 0.13  & 81.18 $\pm$ 0.15& 81.21 $\pm$ 0.17  &  97.33 $\pm$ 0.05 $\uparrow$&  96.97 $\pm$ 0.04& 96.99 $\pm$ 0.03 &80.76 $\pm$ 0.09  & 80.42 $\pm$ 0.08&80.45 $\pm$ 0.07 \\ \midrule
				
				\multicolumn{2}{c|}{Average}   &87.95 $\pm$ 0.16  & 87.69 $\pm$ 0.15 &87.77 $\pm$ 0.15  & 97.45 $\pm$ 0.09 \phantom{$\uparrow$}& 97.18 $\pm$ 0.10 & 97.23 $\pm$ 0.06  &82.21$\pm$ 0.15  & 81.90 $\pm$ 0.15 &81.95 $\pm$ 0.16\\ \bottomrule 
			\end{tabular}
			\label{tab: acc of FedMUA and baseline}
		}
	}
\end{table*}

			\begin{table*} [t]
				\centering 
				\caption{Performance of FedMUA across three datasets in the IID setting.}
				\resizebox{1.01\textwidth}{!}{
					{         
					 \subtable[Under FedEraser setting]{ 	\begin{tabular}{@{}c|c|ccc|ccc|ccc@{}}
							\toprule
							\multirow{2}{*}{\textbf{Values of }} & \multirow{2}{*}{\textbf{Aggregation}} & \multicolumn{3}{c|}{\textbf{Purchase}}& \multicolumn{3}{c|}{\textbf{MNIST}}& \multicolumn{3}{c}{\textbf{CIFAR-10}}  \\ \cmidrule(l){3-11} 
							\textbf{$\epsilon$}	&\textbf{ Rule} & ASR &$Acc_G$(\%)&$\widetilde{Acc}_G$ (\%)& ASR &$Acc_G$(\%)& $\widetilde{Acc}_G$(\%)& ASR &$Acc_G$(\%)& $\widetilde{Acc}_G$ (\%) \\ \midrule
							\multirow{3}{*}{$\epsilon_{max}$} 
							& FedAvg& 0.90&94.09 $\pm$ 0.06  & 94.23 $\pm$ 0.19 & 0.85&97.39 $\pm$ 0.06  & 97.27 $\pm$ 0.07& 0.95&84.20 $\pm$ 0.40  & 83.30$\pm$ 0.26\\
							& Median& 0.90 &94.44 $\pm$ 0.03  &94.51$\pm$ 0.21&0.75&97.40 $\pm$ 0.06& 97.19 $\pm$ 0.10 & 0.85&83.57 $\pm$ 0.55 & 81.43 $\pm$ 0.40 \\
							& Trimmed-mean & 0.90 &94.30 $\pm$ 0.04   & 94.33 $\pm$ 0.22 & 0.80& 97.39 $\pm$ 0.03 &  97.27 $\pm$ 0.09 & 0.95& 84.25 $\pm$ 0.24& 83.32 $\pm$ 0.26  \\ \midrule 
							
							\multirow{3}{*}{$0.8 \times \epsilon_{max}$} 
							& FedAvg& 0.75&94.09 $\pm$ 0.06  & 94.01 $\pm$ 0.17& 0.75&97.39 $\pm$ 0.06  & 97.01 $\pm$ 0.08& 0.85&84.20 $\pm$ 0.40  & 83.42$\pm$ 0.24\\
							& Median& 0.70 &94.44 $\pm$ 0.03  &93.83$\pm$ 0.23&0.70&97.40 $\pm$ 0.06& 96.87 $\pm$ 0.12 & 0.80&83.57 $\pm$ 0.55 & 82.11 $\pm$ 0.35 \\
							& Trimmed-mean & 0.75 &94.30 $\pm$ 0.04   & 94.12 $\pm$ 0.18 & 0.70& 97.39 $\pm$ 0.03 &  97.13 $\pm$ 0.10& 0.80& 84.25 $\pm$ 0.24& 83.42$\pm$ 0.28 \\ \midrule 
							
							\multirow{3}{*}{$0.6 \times \epsilon_{max}$} 
							& FedAvg& 0.65&94.09 $\pm$ 0.06  & 93.95 $\pm$ 0.21 & 0.65&97.39 $\pm$ 0.06  & 96.78 $\pm$ 0.07& 0.80&84.20 $\pm$ 0.40 & 83.52$\pm$ 0.24\\
							& Median& 0.65 &94.44 $\pm$ 0.03  &94.32$\pm$ 0.18&0.60&97.40 $\pm$ 0.06& 96.74 $\pm$ 0.09 & 0.80&83.57 $\pm$ 0.55 & 82.12 $\pm$ 0.38 \\
							& Trimmed-mean & 0.70&94.30 $\pm$ 0.04   & 93.92 $\pm$ 0.21 & 0.60& 97.39 $\pm$ 0.03 &  97.13 $\pm$ 0.08 & 0.75& 84.25 $\pm$ 0.24& 83.52$\pm$ 0.22  \\ \midrule 
							
							\multicolumn{2}{c|}{Average}  &0.77  &94.27 $\pm$ 0.04  & 94.14 $\pm$ 0.20&0.71 &97.39 $\pm$ 0.05  & 97.04 $\pm$ 0.08& 0.84 & 84.00 $\pm$ 0.40 & 82.90 $\pm$ 0.29 \\ \bottomrule 
						\end{tabular}
						\label{tab:FedEraser}
					}
				}
			}
			
			\qquad
					\resizebox{1.01\textwidth}{!}{
				\subtable[Under KNOT setting]{   
						\begin{tabular}{@{}c|c|ccc|ccc|ccc@{}}
						\toprule
						\multirow{2}{*}{\textbf{Values of}} & \multirow{2}{*}{\textbf{Aggregation}} & \multicolumn{3}{c|}{\textbf{Purchase}}& \multicolumn{3}{c|}{\textbf{MNIST}}& \multicolumn{3}{c}{\textbf{CIFAR-10}}  \\ \cmidrule(l){3-11} 
						\textbf{$\epsilon$}	&\textbf{ Rule} & ASR &$Acc_G$(\%)&$\widetilde{Acc}_G$ (\%)& ASR &$Acc_G$(\%)& $\widetilde{Acc}_G$(\%)& ASR &$Acc_G$(\%)& $\widetilde{Acc}_G$ (\%) \\ \midrule
						\multirow{3}{*}{$\epsilon_{max}$} 
						& FedAvg  & 0.95 &81.22 $\pm$ 0.03  & 81.03 $\pm$ 0.01& 0.90&  97.31 $\pm$ 0.02 & 96.79 $\pm$ 0.05 & 0.95 & 80.51 $\pm$ 0.04& 80.61 $\pm$ 0.07\\
						& Median & 0.95 &81.09 $\pm$ 0.04 & 81.39 $\pm$ 0.20 & 0.90& 97.11 $\pm$ 0.05 &97.11 $\pm$ 0.02 & 0.90 & 80.49 $\pm$ 0.06& 80.60 $\pm$ 0.06 \\
						& Trimmed-mean & 0.95 &81.50 $\pm$ 0.04  & 81.04 $\pm$ 0.09 & 0.80 & 97.08 $\pm$ 0.01 &97.14 $\pm$ 0.01 &0.95 & 80.43 $\pm$ 0.03 & 80.57 $\pm$ 0.06\\ \midrule
						
						\multirow{3}{*}{$0.8 \times \epsilon_{max}$} 
						& FedAvg  & 0.90 &81.22 $\pm$ 0.03  & 80.76 $\pm$ 0.02& 0.80&  97.31 $\pm$ 0.02 & 96.55 $\pm$ 0.06 & 0.85 & 80.51 $\pm$ 0.04& 80.41 $\pm$ 0.05\\
						& Median & 0.90 &81.09 $\pm$ 0.04 & 81.13 $\pm$ 0.18 & 0.75& 97.11 $\pm$ 0.05 &96.87 $\pm$ 0.03 & 0.90 & 80.49 $\pm$ 0.06& 80.12 $\pm$ 0.05 \\
						& Trimmed-mean & 0.85 &81.50 $\pm$ 0.04  & 80.85 $\pm$ 0.10& 0.75 & 97.08 $\pm$ 0.01 &97.02 $\pm$ 0.02 &0.85 & 80.43 $\pm$ 0.03 & 80.26 $\pm$ 0.04\\ \midrule
						
						\multirow{3}{*}{$0.6 \times \epsilon_{max}$ } 
						& FedAvg  & 0.85 &81.22 $\pm$ 0.03  & 80.84 $\pm$ 0.02& 0.65&  97.31 $\pm$ 0.02 & 96.94 $\pm$ 0.06 & 0.80& 80.51 $\pm$ 0.04& 80.13 $\pm$ 0.06\\
						& Median & 0.90 &81.09 $\pm$ 0.04 & 81.22 $\pm$ 0.20 & 0.65& 97.11 $\pm$ 0.05 &96.85 $\pm$ 0.03 & 0.80 & 80.49 $\pm$ 0.06& 80.15 $\pm$ 0.05 \\
						& Trimmed-mean & 0.85 &81.50 $\pm$ 0.04  & 80.92 $\pm$ 0.08 & 0.60 & 97.08 $\pm$ 0.01 &96.87 $\pm$ 0.02 &0.75 & 80.43 $\pm$ 0.03 & 80.22 $\pm$ 0.07\\ \midrule
						
						\multicolumn{2}{c|}{Average}  & 0.90 &81.27 $\pm$ 0.37  &81.02 $\pm$ 0.10  & 0.75 &97.17 $\pm$ 0.27&96.90 $\pm$ 0.10 & 0.86 & 80.48 $\pm$ 0.43& 80.34 $\pm$ 0.06 \\ \bottomrule 
					\end{tabular}
					\label{tab:KNOT}
				}
			}
			\end{table*}

	\subsection{Effectiveness of FedMUA} \label{Effectiveness of FedMUA}
	
	We first conduct a comparative study to assess the attack effectiveness of FedMUA. The performance comparison of FedMUA on different datasets, FU methods and aggregation rules in both IID and Non-IID settings are summarized in Table~\ref{tab:iid} and Table~\ref{tab:non-iid}. It's essential to note that the attacker pursues two goals, where ASR-B and ASR assesses Goal I, while $\widetilde{Acc}_G$ and $Acc_G$ evaluate Goal II. 
	
	In the IID setting, experiments with FedMUA involve setting the ratio of malicious unlearning requests to 0.3\%, with the number of malicious clients set to 2. Table~\ref{tab:iid} illustrates that FedMUA achieves an average ASR of 89\%, showcasing its effectiveness across diverse datasets, FU methods, and aggregation rules. We note that the ASR-B is relatively low, FU itself can cause performance degradation to some extent and our attack method can effectively amplify this degradation. Importantly, FedMUA maintains comparable values for $\widetilde{Acc}_G$ and $Acc_G$, indicating minimal degradation in model predictions. However, in KNOT, the values of $\widetilde{Acc}_G$ and $Acc_G$ are generally lower than those in FedEraser, underscoring differences in underlying training mechanisms and strategies.

	Moving to the Non-IID setting, as shown in Table~\ref{tab:non-iid}, FedMUA achieves comparable values for $\widetilde{Acc}_G$ and $Acc_G$. Nevertheless, ASR experiences a slight decrease compared to the IID setting. This reduction could be attributed to the presence of data with diverse feature distributions, which poses a challenge in generalizing across different clients. This diversity makes it more difficult to conduct successful attacks. We also note that the ASR for MNIST is lower than that for CIFAR-10. This can be attributed to two reasons: First, due to its simpler data complexity compared to CIFAR-10, MNIST allows models to more easily learn effective features, enhancing robustness against malicious unlearning attacks. Second, the models used for MNIST are often simpler and have fewer parameters, which can lead to better generalization and increased robustness against such attacks. In addition, FedMUA performs well in non-IID settings mainly because it uses a targeted attack strategy and a feature manipulation technique tailored specifically for the target data. This effectiveness is largely independent of whether the data distribution is IID or non-IID, as our approach concentrates on the characteristics of the target data itself rather than depending on the overall data distribution. In summary, our extensive evaluation underscores the versatility and efficacy of FedMUA in federated unlearning scenarios.

	\subsection{Impact of Ratio of Unlearning Requests} \label{Effectiveness of unlearning samples}
	
	We next compare our proposed FedMUA with the baselines (i.e., \textbf{Rand+MUG}, \textbf{HP} and  \textbf{MSFA}) in terms of ASR under the federated unlearning framework KNOT and FedEraser. In our experiment, we use the FedAvg aggregation rule. The results are depicted in Fig.~\ref{fig:ASR_baseline_FedEraser} and Fig.~\ref{fig:ASR_baseline}, reveal that FedMUA achieves an average ASR of 80\% in FedEraser, with a slight degradation in performance in KNOT. We can also observe that the ASR consistently increases as the ratio of malicious unlearning requests rises under purchase, MNIST, and CIFAR-10 datasets.

	Notably, the ASR of \textbf{Rand+MUG}, \textbf{HP}, and \textbf{MSFA} is significantly lower than that of FedMUA, underscoring the critical role played by the Influential Sample Identification (ISI) component in enhancing the attack effectiveness of FedMUA. It's noteworthy that the ASR of \textbf{Rand+MUG}, \textbf{HP}, and \textbf{MSFA} approaches zero when confronted with the MNIST dataset, and even in the CIFAR-10 dataset, only a 20\% ASR is achieved. Particularly, when dealing with CIFAR-10 dataset in KNOT, the ASR of \textbf{Rand+MUG} slightly increases due to the inherently lower model accuracy. We also observe that the ASR of \textbf{HP} and \textbf{MSFA} is relatively low in most cases. This is primarily because we directly apply \textbf{HP} and \textbf{MSFA} to the federated learning scenario, where multiple clients collaboratively train the model. In such a setting, \textbf{HP} requires a greater amount of poisoned data to be effective, as the distributed nature of federated learning reduce the impact of poisoned data. Similarly, \textbf{MSFA} faces challenges due to the inherent variability and heterogeneity of data across different clients, which can reduce the overall effectiveness of the attack. Additionally, we have tested a higher ratio of malicious unlearning requests (i.e., 10\%) to demonstrate the effectiveness of our method. The experimental results indicate that the ASR of our method can reach 100\% in all scenarios under this setting. The \textbf{Rand+MUG} approach also exhibits a 10\% increase in ASR compared to when there are only 0.1\% malicious unlearning requests. Both \textbf{HP} and \textbf{MSFA} show a slight increase in ASR as well. This emphasizes the inefficacy of randomly selecting samples for malicious unlearning requests during the federated unlearning process, highlighting the superior performance of FedMUA.

	Furthermore, our approach demonstrates resilience, achieving a remarkable average ASR of 75\% even with a low ratio of unlearning requests (e.g., 0.1\%). In the KNOT scenario, the ASR of FedMUA experiences a slight degradation compared to FedEraser, while the ASR of \textbf{Rand+MUG}, \textbf{HP}, and \textbf{MSFA} surpasses the settings in FedEraser. This observation is primarily attributed to the training approach in KNOT, which provides more opportunities for attacks.
	
	Additionally, the performance comparison between FedMUA and the baselines is summarized in Table~\ref{tab: acc of FedMUA and baseline}. We use	$\widetilde{Acc}_{G-B}$ to evaluate the average performance of all the baselines. It's essential to note that $Acc_G$, $\widetilde{Acc}_{G-B}$ and $\widetilde{Acc}_G$ do not fluctuate much when the ratio of unlearning requests varies. $\widetilde{Acc}_{G-B}$ and $\widetilde{Acc}_G$ is slightly lower than $Acc_G$, possibly due to the accuracy degradation caused by unlearning itself. Also, $\widetilde{Acc}_{G-B}$ shows little difference compared to $\widetilde{Acc}_G$, indicating that our attack does not significantly impact accuracy. It is worth noting that when our attack is conducted continuously in a static state, its effect is equivalent to increasing the proportion of unlearning requests. On the other hand, in a real-time scenario, the effectiveness of a continuous FedMUA attack may diminish, as the unlearned data could be removed or the identified influential data may not be sufficiently impactful. Overall, the results affirm that FedMUA exhibits efficient attack performance under various unlearning malicious modifications, highlighting its versatility in navigating different federated learning scenarios.

\begin{figure}[t]
	\centering
	\includegraphics[width=70mm]{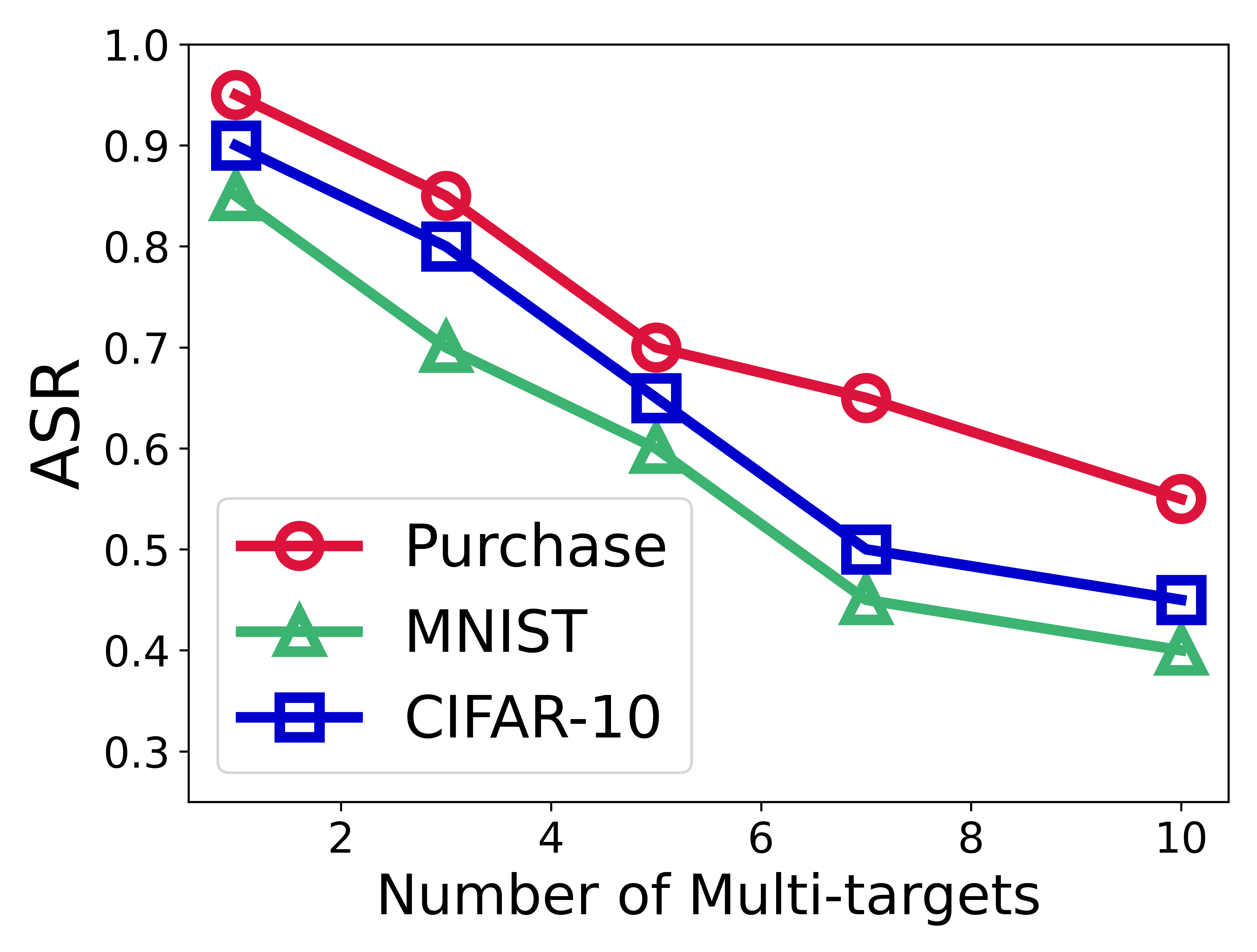}\hspace{2mm}
	\vspace{-5pt}
	\caption{ASR of FedMUA for attacking different numbers of multi-targets.}
	\label{fig:multi-target-varynumber}
\end{figure}

	\begin{figure*}[t]
	\begin{minipage}[t]{0.48\textwidth}
		\centering
		\includegraphics[width=70mm]{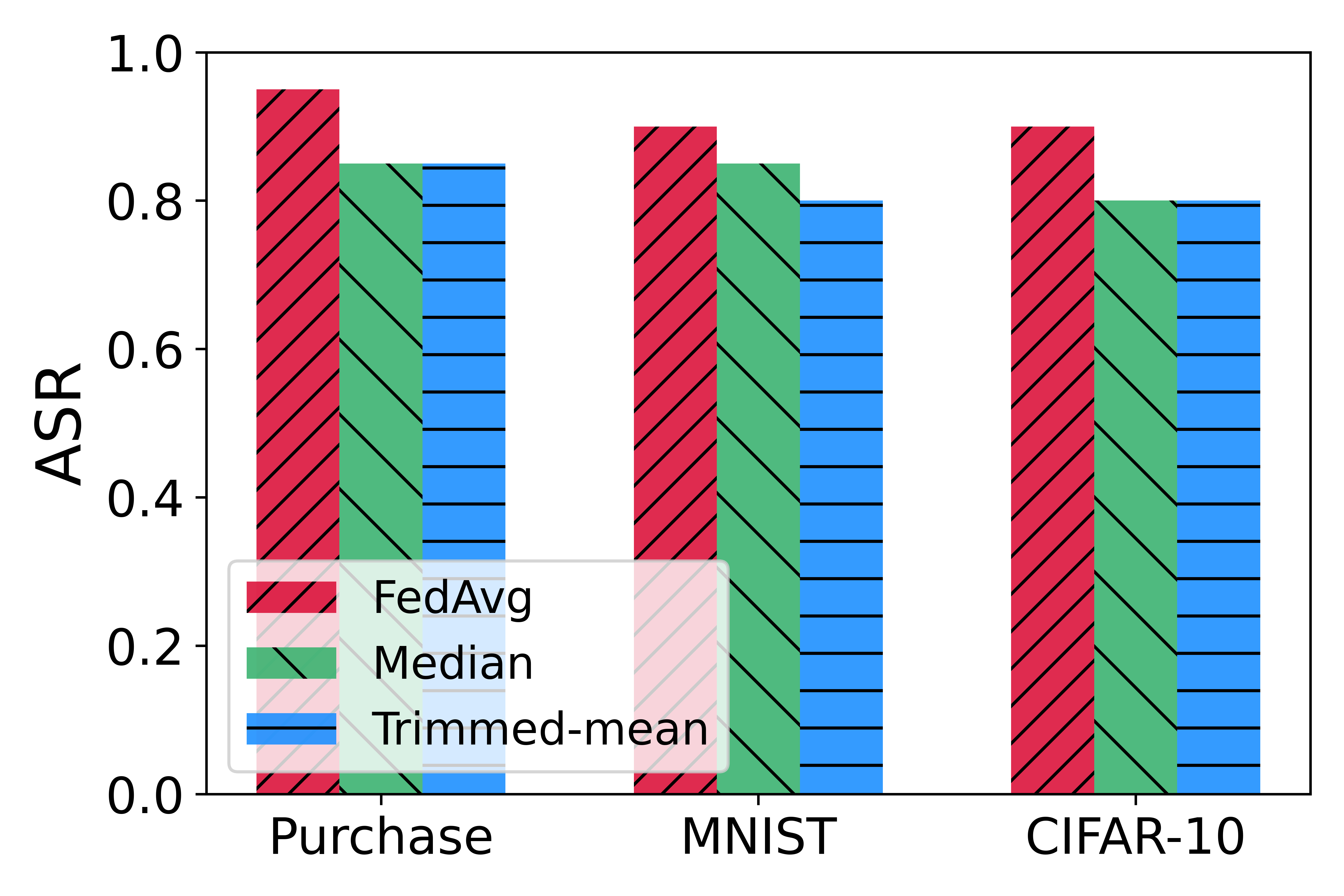}
		\caption{ASR of FedMUA for attacking multi-targets simultaneously.}
		\label{fig:multi-target-fixednumber}
	\end{minipage}
	\hfill
	\begin{minipage}[t]{0.48\textwidth}
		\centering
		\includegraphics[width=70mm]{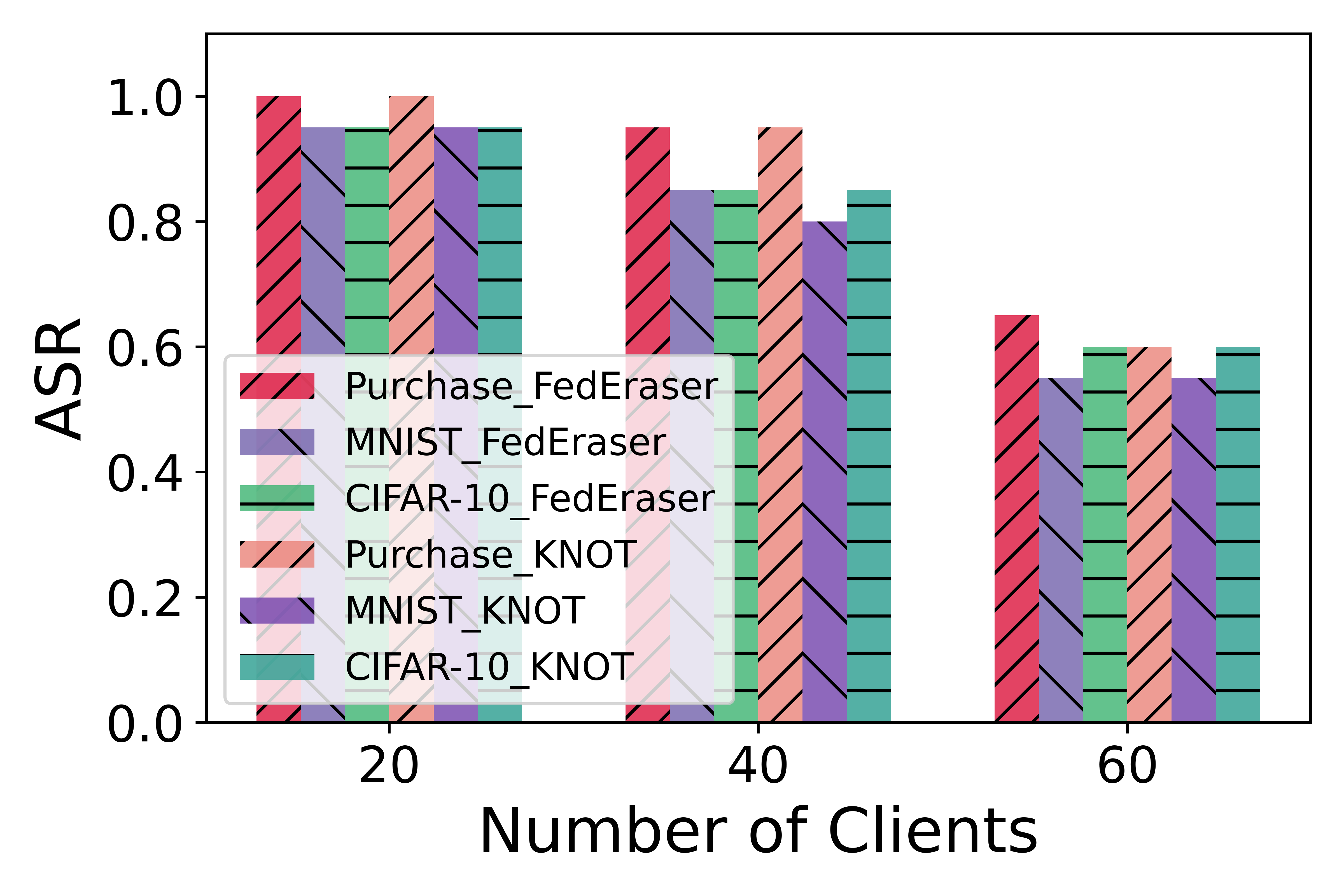}
		\caption{ASR of FedMUA under multi-client scenario.}
		\label{fig:multi-client}
	\end{minipage}
\end{figure*}

\begin{figure*}[t]
	\centering
	\subfigure[Purchase]
	{\includegraphics[width=56mm]{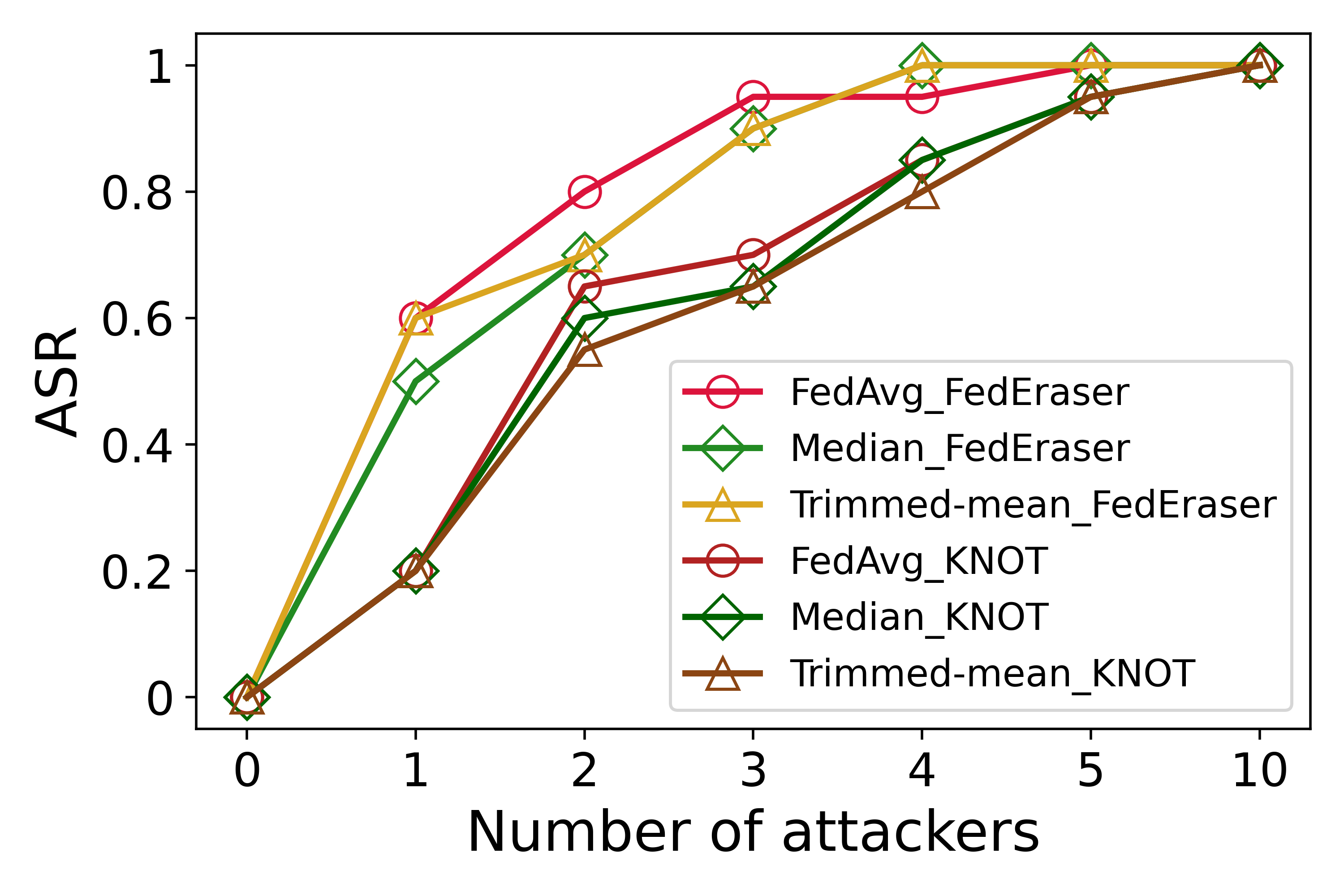}}\label{fig:a}\hspace{2mm}
	\subfigure[MNIST]
	{\includegraphics[width=56mm]{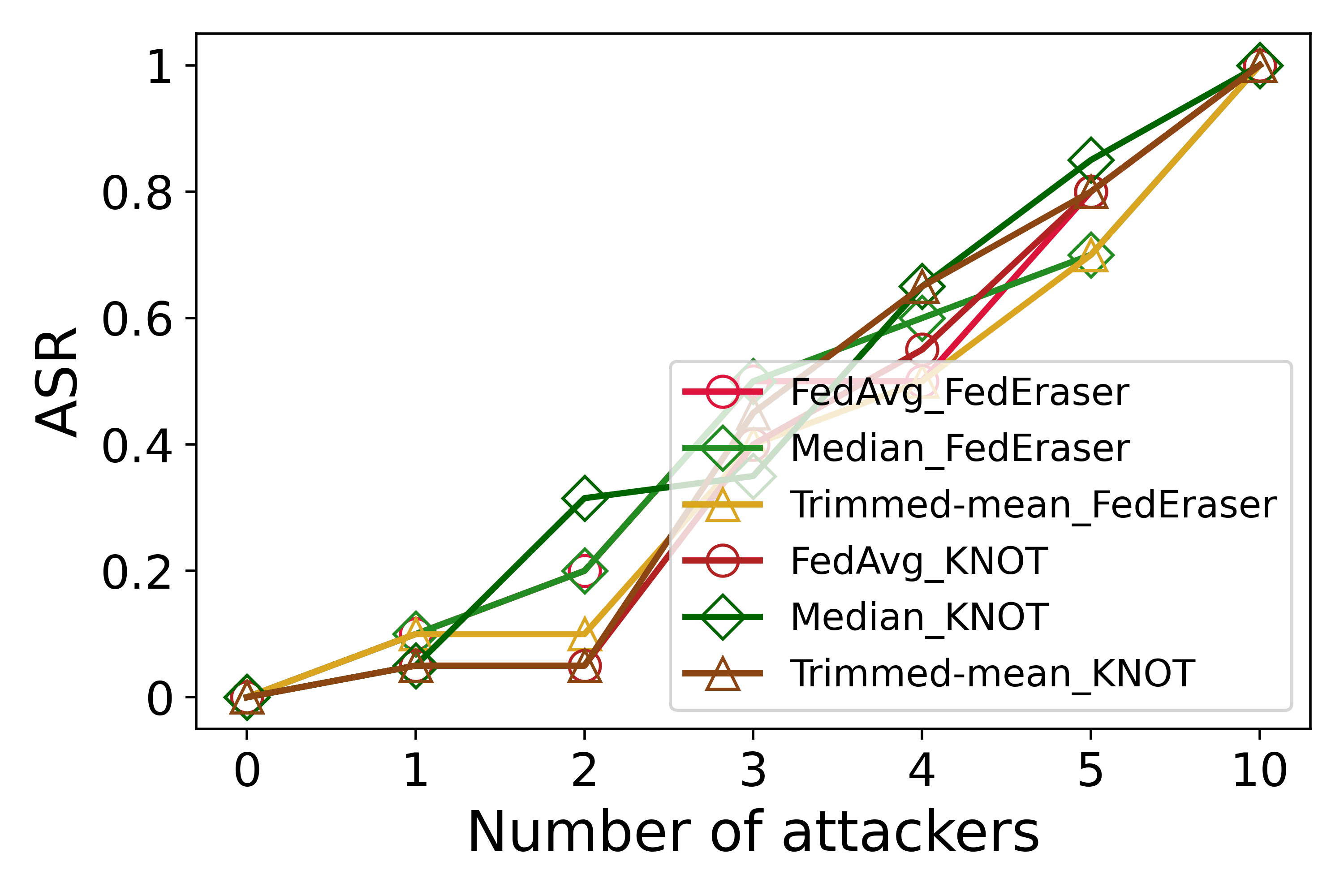}}\label{fig:b}\hspace{2mm}
	\subfigure[CIFAR-10]
	{\includegraphics[width=56mm]{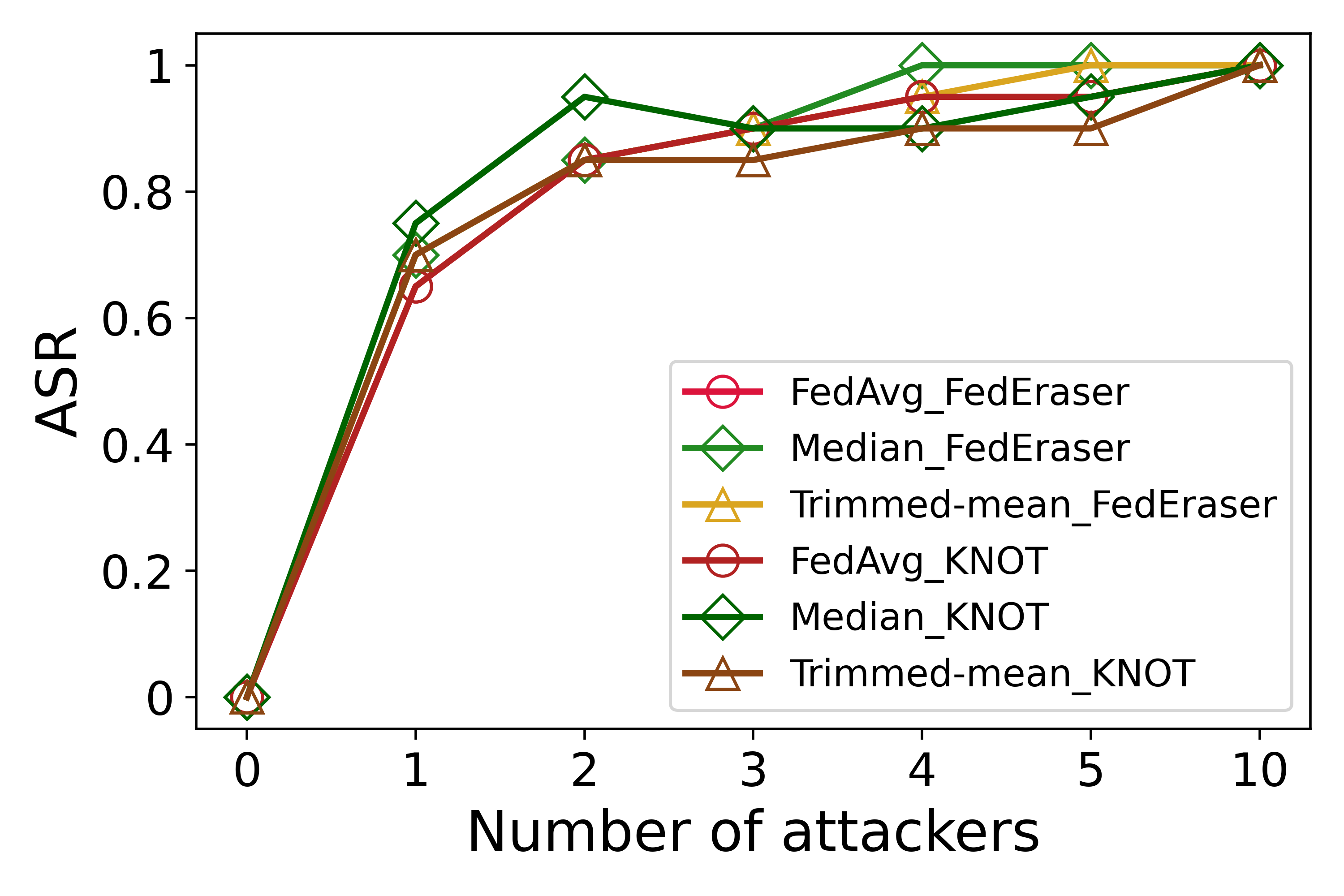}}\label{fig:c}
	\vspace{-5pt}
	\caption{Impact of the number of attackers.}
	\label{fig:impact of number of clients}
	
\end{figure*}

	\subsection{Impact of $\epsilon$} \label{Impact of parameter}
	\label{sec:Hyperparameter}
	In this section, we focus on evaluating the impact of the crucial parameter $\epsilon$ in FedMUA. We conduct experiments to assess how varying values of $\epsilon$ influence the performance of the attack. First, we assess the attack effectiveness on the FedEraser and KNOT in IID setting across varying values of $\epsilon$. Here, $\epsilon$ signifies the manipulated features for influential samples in MUG. A higher $\epsilon$ value will modify more features of influential samples, leading to a more potent attack. The maximum value of $\epsilon$ (i.e., $\epsilon_{max}$) for each influential sample $\boldsymbol x_m$ can be represented as $\lVert \boldsymbol x_m - \boldsymbol x_t \rVert$. In our experiments, we set the ratio of malicious unlearning requests to 0.3\%, with the number of malicious clients set to 2. 
	
	The performance comparison of FedMUA on different datasets, FU methods and aggregation rules in IID settings are summarized in Table~\ref{tab:FedEraser} and Table~\ref{tab:KNOT}. The empirical findings suggest that an increase in $\epsilon$ leads to a rise in ASR. For instance, there is an ASR improvement of approximately 30\%, 20\%, and 30\% for Purchase, MNIST, and CIFAR-10 datasets, respectively when the value of $\epsilon$ increases from $0.6 \times \epsilon_{max}$to $\epsilon_{max}$. Importantly, FedMUA maintains comparable values for $\widetilde{Acc}_G$ and $Acc_G$, indicating minimal degradation in model predictions with changes in $\epsilon$.

\begin{table*} [t]
	\centering 
	\caption{Performance of FedMUA on the real-world dataset in the IID setting.}

	\resizebox{0.9\textwidth}{!}{
		{         
			\begin{tabular}{@{}c|c|cccc|cccc@{}}
				\toprule
				\multirow{2}{*}{\textbf{FU}} & \multirow{2}{*}{\textbf{Aggregation}} & \multicolumn{4}{c|}{\textbf{Credit Score}} & \multicolumn{4}{c}{\textbf{CIFAR-100}} \\ \cmidrule(l){3-10} 
				\textbf{Method}	&\textbf{Rule} & ASR-B & ASR&$Acc_G$(\%)&$\widetilde{Acc}_G$ (\%) & ASR-B & ASR&$Acc_G$(\%)&$\widetilde{Acc}_G$ (\%)\\ \midrule
				\multirow{3}{*}{FedEraser~\cite{liu2021federaser}} 
				& FedAvg  & 0.00& 0.55&57.95 $\pm$ 0.50  & 57.63 $\pm$ 0.61 & 0.00& 0.75&55.67 $\pm$ 0.10  & 55.26 $\pm$ 0.08\\
				& Median & 0.00& 0.60&55.58 $\pm$ 0.27& 55.21 $\pm$ 0.29 & 0.00& 0.80&55.42 $\pm$ 0.12& 55.21 $\pm$ 0.13 \\
				& Trimmed-mean & 0.05& 0.60 &56.71 $\pm$ 0.52 & 55.14 $\pm$ 0.58 & 0.00& 0.80 &55.58 $\pm$ 0.07 & 55.14 $\pm$ 0.09\\ 
				& Krum & 0.00& 0.55&55.39 $\pm$ 0.35 & 55.02 $\pm$ 0.39& 0.00& 0.75&55.03 $\pm$ 0.08 & 54.89 $\pm$ 0.10  \\ \midrule
				
				\multirow{3}{*}{KNOT~\cite{su2023asynchronous} } 
				& FedAvg  & 0.00& 0.50&56.24 $\pm$ 0.46  & 56.01 $\pm$ 0.55 & 0.00& 0.80&53.48 $\pm$ 0.09  & 53.24 $\pm$ 0.08\\
				& Median & 0.00& 0.55&54.37 $\pm$ 0.35& 54.18 $\pm$ 0.34 & 0.05& 0.85&53.76 $\pm$ 0.07& 52.85 $\pm$ 0.28  \\
				& Trimmed-mean & 0.05& 0.55 &55.52 $\pm$ 0.47 & 55.08 $\pm$ 0.49 & 0.00& 0.80 &53.32 $\pm$ 0.02 & 52.97 $\pm$ 0.21\\ 
				& Krum & 0.00& 0.50&54.75 $\pm$ 0.39 & 54.21 $\pm$ 0.34 & 0.00& 0.80&53.15 $\pm$ 0.05 & 52.34 $\pm$ 0.22 \\ \midrule
				
				\multicolumn{2}{c|}{Average}  & 0.012&0.55  &55.81 $\pm$ 0.41  & 55.31 $\pm$ 0.49  & 0.006&0.80  &54.42 $\pm$ 0.08  & 53.98 $\pm$ 0.15 \\ \bottomrule 
			\end{tabular}
			\label{tab:CIFAR100-iid}
		}
	}
\end{table*}

	\begin{figure*}[t]
	\centering
	
	\subfigure[Purchase]
	{\includegraphics[width=58mm]{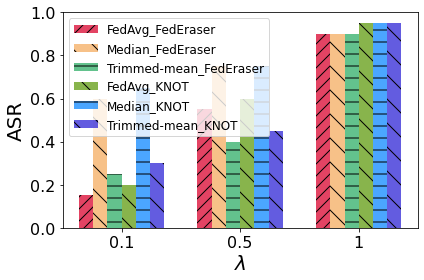}}\label{fig:a}\hspace{2mm}
	\subfigure[MNIST]
	{\includegraphics[width=58mm]{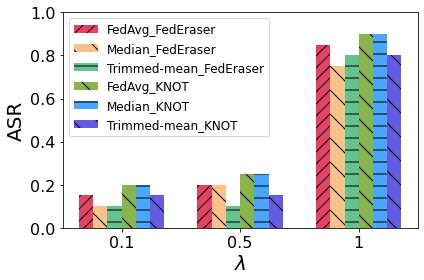}}\label{fig:b}\hspace{2mm}
	\subfigure[CIFAR-10]
	{\includegraphics[width=58mm]{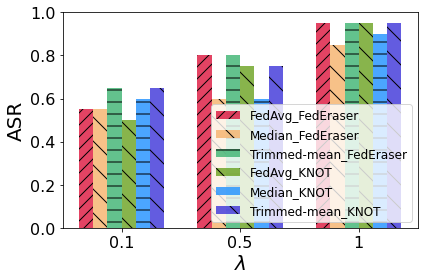}}\label{fig:c}
	\vspace{-5pt}
	
	\caption{Performance of the defense mechanism.}
	\label{fig:Performance of the defense mechanism}
\end{figure*}

			\begin{figure}[t]
	\centering	
	\subfigure[FAT]{\label{defense_FAT}  \includegraphics[scale=0.28]{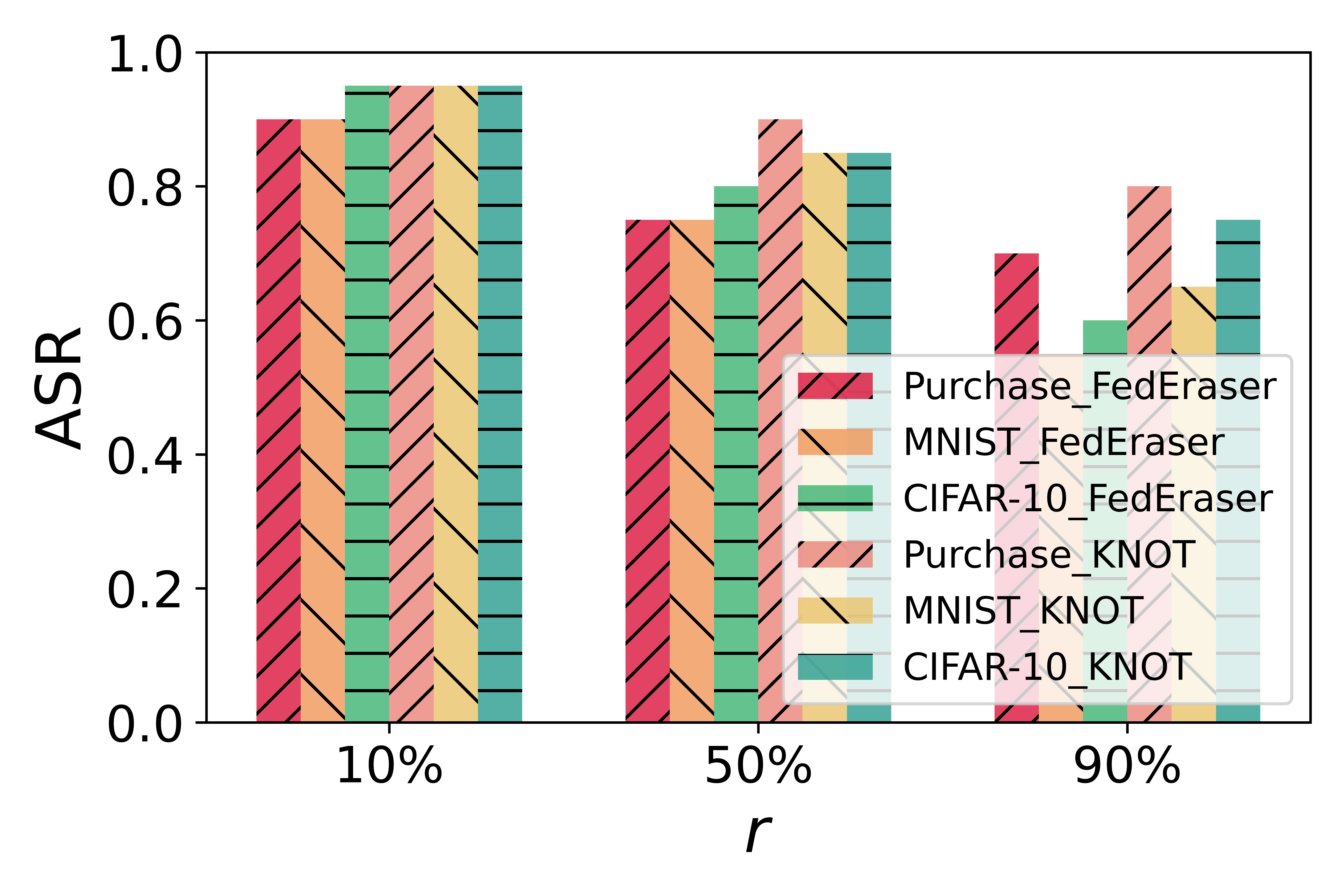}}
	\subfigure[FADngs]{\label{defense_FADngs} \includegraphics[scale=0.28]{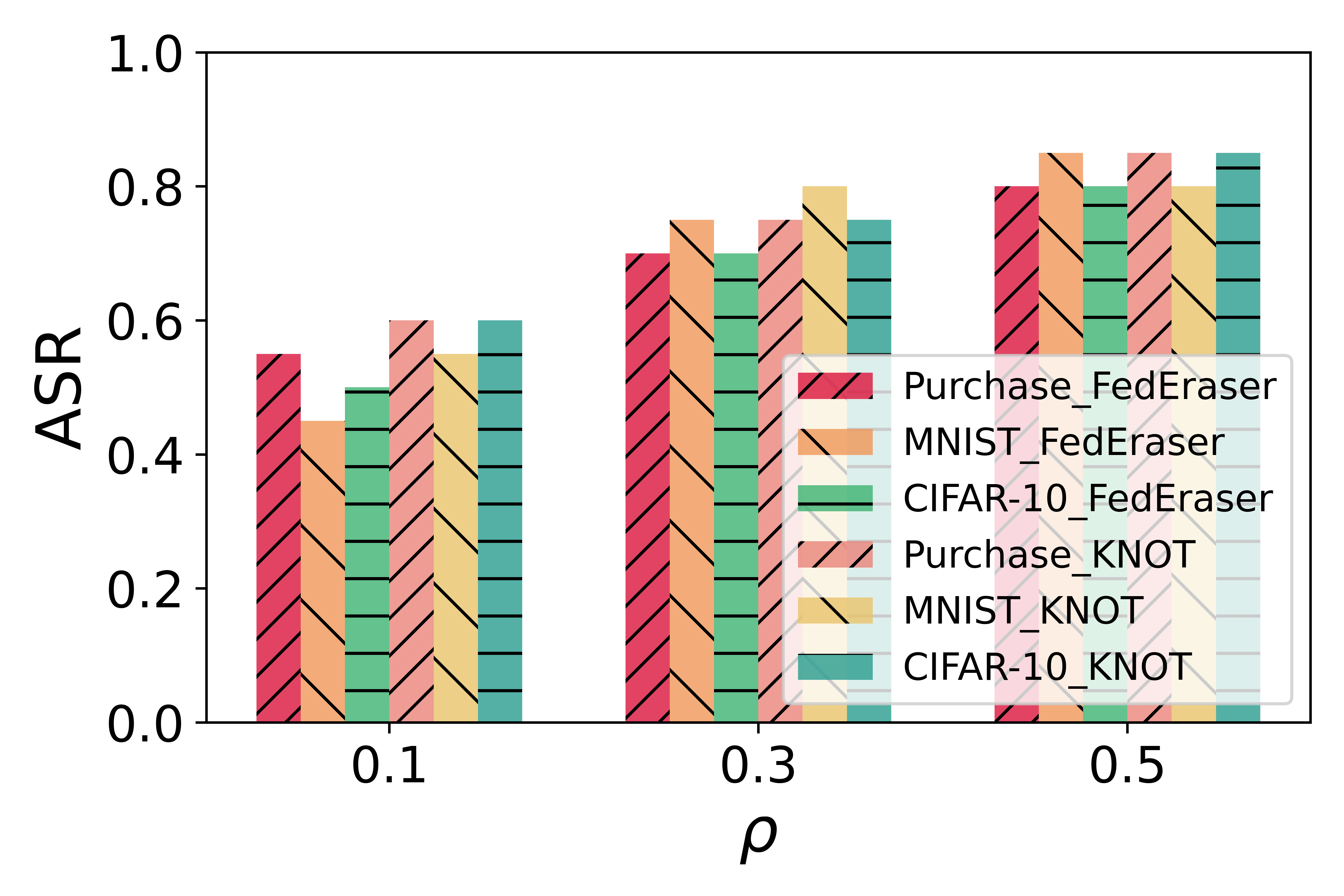}}
	
	\caption{ASR of baseline defenses against FedMUA.}
	\label{fig:baseline_defense}
\end{figure}

	\subsection{Multi-Target Scenario} \label{Multi-target Mode}
	\label{sec:Hyperparameter}
	Then, we extend our investigation to a more realistic setting where the target objects are multiple. We specifically explore and experiment with scenarios involving multiple targets to provide a more comprehensive understanding of FedMUA in practical and diverse federated learning contexts. In this section, we assess the effectiveness of FedMUA in simultaneously attacking multiple targets across the three datasets. In this scenario, we initiate the process by identifying influential samples for each target and then selecting the union of influential samples to induce deviations across all targets. The MUG part employs the same methodology as for attacking a single target. In our experiments, we initially set the ratio of influential samples to 0.3\% and evaluate the effectiveness of FedMUA in FedAvg. The results presented in Fig.~\ref{fig:multi-target-varynumber} indicate a slight decrease in ASR for the three datasets as the number of targets increases from 1 to 9. Overall, there is an approximate 45\% decrease in ASR on average.
	
	Additionally, we conduct simultaneous attacks on three targets for the three datasets and evaluate the effectiveness of FedMUA under different aggregation rules. The results in Fig.~\ref{fig:multi-target-fixednumber} show that the ASR of FedMUA keeps a relatively high value under different aggregation rules. Overall, these findings highlight the adaptability and impact of FedMUA in scenarios where multiple targets are concurrently subjected to the attack. The observed changes in ASR shed light on the robustness and potential implications of such multi-target attacks in federated learning settings.

		\subsection{Multi-Client Scenario} \label{Performance on Multi-Client Scenario}
		In this section, we evaluate the performance in a multi-client scenario. We conduct experiments using the Purchase, MNIST, and CIFAR-10 datasets with client numbers set at 20, 30, 40, and 50. The number of malicious clients is fixed at 2, and the ratio of malicious unlearning requests is set to 1\%. As shown in Fig.~\ref{fig:multi-client}, the ASR remains high when the number of participants is 20 or fewer. However, the ASR drops significantly when the number of participants reaches 60, which is reasonable since the attacker controls only 3\% of the clients uploaded to the server. As the number of participants increases, the ASR continues to decrease due to the dilution of the attacker's influence among a larger number of clients.

	\subsection{Impact of the Number of Attackers } \label{Impact of number of attackers }
	In this section, we discuss the impact of the number of attackers (i.e., malicious clients) on ASR.  In our experiments, we set the total number of clients to $20$. Fig.~\ref{fig:impact of number of clients} illustrates the attack performance of FedMUA when a different number of clients send malicious unlearning requests to the server. 
	
	In general, more attackers result in a higher ASR. It is evident that having more malicious unlearning requests increases the likelihood that the global model will learn the malicious characteristics of the unlearning samples, thereby contributing to a higher ASR on the target sample. For instance, when there is only 1 attacker, we notice that the ASR is higher than 60\% in CIFAR-10 but only reaches 10\% in MNIST. This disparity is due to the inherent complexity of CIFAR-10, which has higher variability and more intricate features compared to MNIST. The complexity of CIFAR-10 provides more opportunities for attackers to exploit vulnerabilities in the model, making it easier to insert malicious characteristics into the unlearning process. In contrast, MNIST's simpler and more homogeneous nature makes it harder for a single attacker to significantly influence the global model, resulting in a much lower ASR. As the number of attackers increases, the ASR correspondingly rises. Specifically, when the number of attackers is increased to 4, the ASR can achieve approximately 80\% on average in most cases. Additionally, we tested a higher proportion of malicious clients, specifically 50\%, to demonstrate the robustness of our method. The experimental results indicate that the ASR can reach 100\% in all scenarios under this setting. This trend highlights the critical impact that the number of malicious clients can have on the effectiveness of unlearning attacks and underscores the importance of developing robust defense mechanisms to mitigate such threats.

	\subsection{Performance on the Real-World Dataset} \label{Performance on Real-World dataset}
		In this section, we assess the effectiveness of FedMUA on the more practical dataset (i.e., Credit Score and CIFAR-100) under the IID setting. We configure the experiment with a malicious unlearning request ratio of 0.3\% and set the number of malicious clients to 2. The results of FedMUA's performance are summarized in Table~\ref{tab:CIFAR100-iid}. We can observe that the ASR-B is close to 0, while our attack method achieves an average ASR of 55\% for the Credit Score dataset and 80\% for the CIFAR-100 dataset. This demonstrates the robustness and effectiveness of FedMUA, even when applied to the more practical datasets. The Credit Score dataset is less susceptible to adversarial manipulation primarily because it contains numerical and categorical features, which are harder to manipulate compared to image data like CIFAR-100. This makes it more challenging for the attack to significantly alter the model's predictions. Additionally, FedMUA maintains comparable values for $\widetilde{Acc}_G$ and $Acc_G$, indicating minimal degradation in model predictions. This suggests that FedMUA not only effectively handles malicious unlearning requests but also preserves the overall integrity and accuracy of the model.

	\begin{table}[t]
		\centering
		\caption{Time overhead for FedMUA's components (s).} \label{tab:time overhead}
		{\fontsize{8.5}{11.5}\selectfont
			\begin{tabular}{|c|c|c|c|l|}
				\hline
				\diagbox{Component}{Time overhead}{Dataset}                 & Purchase                  & MNIST & \multicolumn{2}{c|}{CIFAR-10 } \\ \hline
				\multicolumn{1}{|c|}{ISI} & 16.273                   & 25.504     & \multicolumn{2}{c|}{62.535}          \\ \hline
				\multicolumn{1}{|c|}{MUG}  & 0.091                 & 0.104     & \multicolumn{2}{c|}{0.106}         \\ \hline
			\end{tabular}
		}
	\end{table}

	\subsection{Performance of the Defense Mechanism } \label{Performance of the defense mechanism}
	In this section, we evaluate the effectiveness of our proposed defense mechanism against malicious unlearning attacks during the unlearning process. We use the same parameter settings as in the experiments on malicious unlearning attacks in the IID setting. Specifically, we set the ratio of malicious unlearning requests to 0.3\% and the number of malicious clients set to 2. We compare the ASR before and after the defense. Fig.~\ref{fig:Performance of the defense mechanism} presents the performance of the proposed defense mechanism against FedMUA for various choices of $\lambda$. When we set $\lambda=1$, it means the defender did not alter the value of the malicious gradients, and the ASR should remain the same as in FedMUA.  For smaller values of $\lambda$, more of the malicious gradient values are reduced, and the ASR decreases further. This indicates that the defense performance of the proposed mechanism is better. In particular, when $\lambda=0.1$, the average ASR for Purchase, MNIST, and CIFAR-10 is 35\%, 15\%, and 58\%, respectively. The average ASR has been reduced by 50\% compared to the scenario when $\lambda=0.1$. In general, the ASR in CIFAR-10 is higher than in the other two datasets. This is mainly because CIFAR-10 is more complex, and the defense mechanism demonstrates lower defense robustness. For different FU methods and aggregation rules, we can observe that KNOT has higher attack robustness compared to FedEraser. This is because KNOT employs more sophisticated techniques for gradient updates, ensuring a more resilient model training process. It's also important to note that our defense approach, which mainly involves comparing and reducing the magnitudes of update gradients, requires relatively low computational resources. As a result, it does not significantly impact the system's overall performance.
	
	In addition, Fig.~\ref{defense_FAT} illustrates the performance of the FAT defense against FedMUA. We report the ASR as the ratio of adversarial samples varies. The results are averaged across FedAvg, Median, and Trimmed-mean methods. It is evident that FAT is not effective in significantly reducing the ASR, with the average ASR still reaching 65\% even when 90\% of the samples are adversarial. Fig.~\ref{defense_FADngs} shows the ASR for the FADngs defense against FedMUA. By adjusting the shrinkage intensity $\rho$ to different levels, we observe that although FADngs performs better than FAT, it only reduces the ASR by an average of 20\%. While adjusting the shrinkage intensity can help in tuning the defense, it may not be sufficient to fully counteract the impact of unlearning attacks and might not effectively adapt to their patterns.

			\subsection{Evaluation on Run-time Overhead} \label{Time Overhead}
			Lastly, we evaluate the run-time overhead of FedMUA, which was tested on an NVIDIA Geforce RTX 4090. The experimental results are detailed in Table~\ref{tab:time overhead}. The run-time for FedMUA is divided into two parts: ISI and MUG.
			
			In the ISI part, we assess the time required to calculate $100$ influential samples. Notably, FedMUA takes the longest time with the CIFAR-10 dataset, requiring $62.535$ seconds to compute these samples. In contrast, the Purchase dataset demands less time than CIFAR-10. This discrepancy is primarily due to the dataset complexity, as larger datasets like CIFAR-10 naturally require more computational resources and time for processing.
			
			In the MUG part, we measure the time taken to generate $100$ malicious unlearning samples. For the Purchase, MNIST, and CIFAR-10 datasets, FedMUA requires only $0.091$ seconds, $0.104$ seconds, and $0.106$ seconds, respectively. The significantly reduced time in the MUG phase is largely attributed to the nature of the task, which involves merely altering the features of the data. This process is inherently faster because it does not require the extensive computations involved in identifying influential samples. Additionally, the efficiency in the MUG phase highlights the optimization of FedMUA in handling feature modifications fastly, further emphasizing its suitability for scenarios where rapid unlearning is crucial.

			\section{Conclusion}
			
			\label{sec:Conclusion}
			
			In this work, we have introduced FedMUA, the first framework designed to expose potential threats within the unlearning process of federated learning models. To make our attack more efficient, we devised a novel two-step attack method, namely ISI and MUG, aimed at identifying and crafting malicious unlearning requests. Furthermore, to mitigate such unlearning risks, we also propose a rapid and effective defense mechanism based on the observation that gradient updates from malicious clients are typically larger than those from clean clients. Extensive evaluations across three public datasets substantiate FedMUA's capacity to effectively launch attacks, inducing manipulated predictions for the target data. We envisage that our work will prompt the FL community to pay closer attention to the security implications inherent in existing federated unlearning techniques.
		
{
	\balance
	\bibliographystyle{IEEEtran}
	\bibliography{mybib}
}

	\end{document}